\RequirePackage{lineno}
\documentclass[prd,twocolumn,amsmath,amssymb]{revtex4-2}
\usepackage{graphicx}
\usepackage{dcolumn}
\usepackage{bm}
\usepackage{epsfig}
\usepackage{overpic}
\usepackage{verbatim}
\usepackage{xspace}
\usepackage{booktabs}
\usepackage{cuted}
\usepackage{helvet}
\usepackage{sansmath}
\newsavebox{\tablebox}
\usepackage[colorlinks,linkcolor=blue,anchorcolor=red,citecolor=blue,breaklinks=true]{hyperref}
\usepackage{subfig}
\usepackage{cleveref}

\usepackage{rotating}
\graphicspath{{figures/}}

\newcommand{\Rmnum}[1]{\uppercase\expandafter{\romannumeral #1}}
\newcommand{\BR}{{\cal B}}

\newcommand{\EE}{e^+e^-}

\def\gevcc{\ifmmode {\mathrm{Ge\kern -0.1em V}/c^2}\else
                   {\textrm{Ge\kern -0.1em V}/$c^2$}\fi}%
\def\gev{\ifmmode {\mathrm{Ge\kern -0.1em V}}\else
                   {\textrm{Ge\kern -0.1em V}}\fi}%
\def\mevcc{\ifmmode {\mathrm{Me\kern -0.1em V}/c^2}\else
                   {\textrm{Me\kern -0.1em V}/$c^2$}\fi}%
\def\mev{\ifmmode {\mathrm{Me\kern -0.1em V}}\else
                   {\textrm{Me\kern -0.1em V}}\fi}%
\newcommand{\pip}{\pi^+}
\newcommand{\pim}{\pi^-}

\newcommand{\bfg}{\begin{figure}}
\newcommand{\efg}{\end{figure}}
\newcommand{\bitm}{\begin{itemize}}
\newcommand{\eitm}{\end{itemize}}
\newcommand{\bnum}{\begin{enumerate}}
\newcommand{\enum}{\end{enumerate}}
\newcommand{\btbl}{\begin{table}}
\newcommand{\etbl}{\end{table}}
\newcommand{\btbu}{\begin{tabular}}
\newcommand{\etbu}{\end{tabular}}
\newcommand{\beq}{\begin{equation}}
\newcommand{\edq}{\end{equation}}
\def\rfpi{r^{4\pi}_{\bf{p}}}
\def\sp{\Delta\delta_{D}}
\def\spp{\Delta\delta_{\bf{p}}}

\def\fpi{\pi^{+}\pi^{-}\pi^{+}\pi^{-}}
\def\Fpi{4\pi}
\def\Nst{N_{\textrm{ST}}}
\def\Ndt{N_{\textrm{DT}}}
\def\dE{\Delta E}
\def\mbc{m_{\textrm{BC}}}
\def\trm{\textrm}
\def\missm{M^2_{\textrm{miss}}}
\def\umiss{U_{\textrm{miss}}}

\newcommand{\ks}{K_S^0}
\newcommand{\kl}{K_L^0}
\newcommand{\ksl}{K_{S,L}^0}
\newcommand{\pipi}{\pi^{+}\pi^{-}}

\newcommand{\Dz}{D^{0}}
\newcommand{\Dzbar}{\bar{D}^{0}}

\newcommand{\psip}{\psi(3686)}
\newcommand{\psipp}{\psi(3770)}
\newcommand{\jpsi}{J/\psi}

\def\piz{\pi^0}
\def\pip{\pi^+}
\def\pim{\pi^-}
\def\CP{C\!P}

\begin{document}
\normalsize
\parskip=5pt plus 1pt minus 1pt

\title{\boldmath Model-independent determination of the strong-phase difference \\ between  $\Dz$ and $\Dzbar\to\fpi$ decays}

\author{
M.~Ablikim$^{1}$, M.~N.~Achasov$^{4,c}$, P.~Adlarson$^{76}$, O.~Afedulidis$^{3}$, X.~C.~Ai$^{81}$, R.~Aliberti$^{35}$, A.~Amoroso$^{75A,75C}$, Q.~An$^{72,58,a}$, Y.~Bai$^{57}$, O.~Bakina$^{36}$
% ...
\\
\vspace{0.2cm}
(BESIII Collaboration)\\
\vspace{0.2cm} {\it
$^{1}$ Institute of High Energy Physics, Beijing 100049, People's Republic of China\\
% ...
$^{3}$ Bochum  Ruhr-University, D-44780 Bochum, Germany\\
$^{4}$ Budker Institute of Nuclear Physics SB RAS (BINP), Novosibirsk 630090, Russia\\
% ...
$^{35}$ Johannes Gutenberg University of Mainz, Johann-Joachim-Becher-Weg 45, D-55099 Mainz, Germany\\
$^{36}$ Joint Institute for Nuclear Research, 141980 Dubna, Moscow region, Russia\\
% ...
$^{57}$ Southeast University, Nanjing 211100, People's Republic of China\\
$^{58}$ State Key Laboratory of Particle Detection and Electronics, Beijing 100049, Hefei 230026, People's Republic of China\\
% ...
$^{72}$ University of Science and Technology of China, Hefei 230026, People's Republic of China\\
% ...
$^{75}$ University of Turin and INFN, (A)University of Turin, I-10125, Turin, Italy; (B)University of Eastern Piedmont, I-15121, Alessandria, Italy; (C)INFN, I-10125, Turin, Italy\\
$^{76}$ Uppsala University, Box 516, SE-75120 Uppsala, Sweden\\
% ...
$^{81}$ Zhengzhou University, Zhengzhou 450001, People's Republic of China\\
\vspace{0.2cm}
$^{a}$ Deceased\\
% ...
$^{c}$ Also at the Novosibirsk State University, Novosibirsk, 630090, Russia\\
% ...
}
}

\begin{abstract}

Measurements of the strong-phase difference between $D^0$ and $\Dzbar\to\fpi$ are performed in bins of phase space.  The study exploits a sample of quantum-correlated $D\bar{D}$ mesons collected by the BESIII experiment in $e^+e^-$ collisions at a center-of-mass energy of 3.773~GeV, corresponding to an integrated luminosity of 2.93~fb$^{-1}$. Here, $D$ denotes a neutral charm meson in a superposition of flavor eigenstates.
The reported results are valuable for measurements of the $C\!P$-violating phase $\gamma$ (also denoted $\phi_3$) in $B^\pm \to DK^\pm$, $D \to \fpi$ decays, and the binning schemes are designed to provide good statistical sensitivity to this parameter. The expected uncertainty on $\gamma$ arising from the precision of the strong-phase measurements, when applied to very large samples of $B$-meson decays, is around $1.5^\circ$ or $2^\circ$, depending on the binning scheme.  The binned strong-phase parameters are combined to give a value of $F_+^{4\pi} = 0.746 \pm 0.010 \pm 0.004$ for the $C\!P$-even fraction of $D^0 \to \fpi$ decays, which is around 30\% more precise than the previous best measurement of this quantity.
\end{abstract}


\maketitle

\flushbottom

\section{Introduction}

In the Standard Model of particle physics, $C\!P$ violation is accommodated through a single complex phase in the Cabibbo-Kobayashi-Maskawa (CKM) matrix, which relates the weak and mass eigenstates in the quark sector~\cite{PhysRevLett.10.531,10.1143/PTP.49.652}.   Studies of $C\!P$ violation are often performed in the context of the Unitarity Triangle, which is a geometrical representation of the CKM matrix in the complex plane.  One parameter of this triangle, the angle $\gamma$ (sometimes denoted $\phi_3$) $= \arg(-V_{us}{V_{ub}}^\ast / V_{cs} {V_{cb}}^*)$ holds particular importance, as it can be determined directly in tree-level $B^\pm \to DK^\pm$ decays, where $D$ is a superposition of $D^0$ and $\bar{D^0}$ flavor-eigenstates, and indirectly from processes that involve virtual loops.  A comparison of the results from the two methods is a powerful tool to probe for physics effects beyond the Standard Model, as these are expected to manifest themselves more strongly in the loop-driven processes.   The sensitivity of this comparison is currently limited by the tree-level measurement of $\gamma$, which is significantly less precise than the one from the loop-level ~\cite{PhysRevD.84.033005,Bona:2005vz}.  The tree-level measurement is dominated by analyses involving decays such as $D \to K^\mp \pi^\pm$, $D \to K^+K^-$, $D \to K^0_S \pi^+\pi^-$ and $D \to K^\mp \pi^\mp \pi^+\pi^-$~\cite{LHCb:2020hdx,LHCb:2020yot,LHCb:2022nng, Belle-II:2024eob}. As noted in Ref.~\cite{Harnew:2017tlp}, the decay $D \to \fpi$ is a promising addition to this ensemble of modes, but its inclusion necessitates having good knowledge of certain parameters that characterize the decay, as described below. These parameters are best accessed in charm-threshold experiments such as BESIII.

$C\!P$ violation can be studied in $B^\pm \to DK^\pm$, $D\to \fpi$ decays with two strategies.   In the first approach, a $C\!P$ asymmetry is measured between the total $B^+$ and $B^-$ decay rates, which can be expressed in terms of $\gamma$ and other parameters related to the $B$-meson decay provided that the $C\!P$-even fraction $F^{4\pi}_+$ of  $D^0 \to \fpi$ decays is known~\cite{Malde:2015mha}.  BESIII recently reported $F^{4\pi}_+ = 0.735 \pm 0.015 \pm 0.005$~\cite{BESIII:2022wqs}, which enables such an interpretation, and indicates that this decay mode is predominantly $C\!P$ even.   
The second approach, which can achieve greater sensitivity to $\gamma$, involves constructing asymmetries in suitably chosen bins of the phase space of the $D$-meson decay. This method is analogous to the one proposed and successfully exploited for the analysis of $B^\pm \to DK^\pm$, $D \to K^0_S \pi^+\pi^-$ decays~\cite{bondar:proceedings,GGSZ,PoluektovFeasibility,PoluektovDPairs,PhysRevD.82.112006,LHCb:2020yot}. In this approach, it is necessary to know the parameters that characterize the $D$ decay. The most important parameters are those associated with the difference in strong phase between the $D^0$ and $\bar{D^0}$ decays in each bin of phase space.

This paper reports measurements of the strong-phase parameters in $D \to \fpi$ decays, based on a data set of $e^+e^- \to \psi(3770) \to D \bar{D}$ events collected by the BESIII experiment, corresponding to an integrated luminosity of 2.93~${\rm fb}^{-1}$.  
The strong-phase parameters are measured in bins of phase space. These results are also used to determine an updated value of $F^{4\pi}_+$ for the decay, which supersedes the previous measurement performed on the identical data set~\cite{BESIII:2022wqs}.
The binning schemes used are guided by a recent amplitude model constructed from the same sample~\cite{BESIII:2023exz}. 
The results complement previous $\textrm{BESIII}$ measurements of strong-phase parameters in bins of phase space for other multi-body $D$ decays~\cite{BESIII:2020hlg,BESIII:2020khq, BESIII:KsKK, Ablikim:2021cqw}.  The data set is around four times larger than the one collected by CLEO-c, which was used to make a first measurement of localized strong-phase parameters in $D \to \fpi$ decays~\cite{Harnew:2017tlp}.  The larger sample allows for more sensitive binning schemes to be devised, and for higher statistical sensitivity to be achieved for the strong-phase parameters.

\section{Formalism and measurement strategy}
\label{sec:formalism}

Neutral $D$ mesons produced in $e^+e^-$ collisions at the $\psipp$ resonance have no accompanying particles.  Therefore, their subsequent decays are quantum correlated and retain the $C$-odd eigenvalue of the initial state.
When one charm meson decays to the signal mode $S$ in the phase-space region $i$, and the other decays to a tag mode $G$ in the phase-space region $j$, the decay width of $\psipp \to D\bar{D} \to S_{\rm{i}}G_{\rm{j}}$ is given by
\begin{widetext}
\begin{equation}
\label{eq:DDfigj}
	\Gamma[\psi(3770)\to D\bar{D}\to S_{\rm{i}}G_{\rm{j}}]\propto [T^S_i\bar{T}^G_j +\bar{T}^S_iT^G_j - 2\sqrt{T^S_i\bar{T}^G_j\bar{T}^S_iT^G_j}(c^S_ic^G_j+ s^S_is^G_j)], 
\end{equation}
\end{widetext}
where $T^{S(G)}_i$ is the fraction of $\Dz\to S(G)$ decays in phase-space region $i$, and $\bar{T}^{S(G)}_i$ is the analogous quantity for $\Dzbar$ decays. The parameters $c_i$ and $s_i$ are the amplitude-weighted average sine and cosine of the strong-phase difference between $\Dz$ and $\Dzbar$ decays, given by 
\begin{linenomath*}
\begin{equation}
\label{eq:CiSi}
\begin{split}
c_i  & \equiv \frac{\int_i|A_\textbf{p}| |\bar{A}_\textbf{p}| \mathrm{cos}(\spp) \mathrm{d}\textbf{p} }{\sqrt{\int_i |A_\textbf{p}|^2\mathrm{d}\textbf{p}\int_i |\bar{A}_\textbf{p}|^2\mathrm{d}\textbf{p}}}, \\
s_i  & \equiv \frac{\int_i|A_\textbf{p}| |\bar{A}_\textbf{p}| \mathrm{sin}(\spp) \mathrm{d}\textbf{p} }{\sqrt{\int_i |A_\textbf{p}|^2\mathrm{d}\textbf{p}\int_i |\bar{A}_\textbf{p}|^2\mathrm{d}\textbf{p}}}, 
\end{split}
\end{equation}
\end{linenomath*}
where $A_\textbf{p}$ ($\bar{A}_\textbf{p}$) is the decay amplitude of the $\Dz$ ($\Dzbar$) meson at the phase-space point $\textbf{p}$, $\spp = \trm{arg}(A_\textbf{p}) - \trm{arg}(\bar{A}_\textbf{p})$ is the strong-phase difference and the integral $\int_i \mathrm{d}\textbf{p}$ is performed over region $i$ of phase space. Equation~(\ref{eq:DDfigj}) holds under the assumption of no $\CP$ violation in the charm-meson decays and  omits terms of ${\cal O}(x^2,y^2)\sim 10^{-5}$ associated with $D^0$-$\bar{D}^0$ oscillations, where $x$ and $y$ are the usual mixing parameters~\cite{pdg}. Both assumptions are valid within the expected precision of this measurement. The phase-space regions are numbered from $-i$ to $i$, with no bin $0$. The bin $+i$ maps to bin $-i$ under $\CP$ transformation. It follows that $c_i=c_{-i}$, $s_{i}=-s_{-i}$ and $\bar{T}_i = T_{-i}$. 

In order to determine the $c_i$ and $s_i$ parameters, two types of tags are used. The first are $D$ meson decays to $\CP$ eigenstates or $D \to \pip\pim\piz$, which is very close to being a $\CP$-even eigenstate with $\CP$-even fraction $F_+^{\pi\pi\pi^0}= 0.973 \pm 0.017$~\cite{Malde:2015mha}. For these tags, the decay phase space is treated inclusively, meaning $s_i$ is 0, and $c_i$ is given by $(2F_{+}-1)$. This results in $c_i$ being $-1$ for a $\CP$-odd eigenstate with $F_+ = 0$, and $+1$ for a $\CP$-even eigenstate with $F_+=1$.
Hence, the observed double-tag (DT) yield in events where one $D$ meson decays to signal and the other to the $\CP$ eigenstate is given by
\begin{linenomath*}
\begin{equation}
\label{eq:DTCP}
M_i = h_{\CP}\left(T^S_i+T^S_{-i}-2c_i(2F_+-1)\sqrt{T^S_iT^S_{-i}}\right),
\end{equation}
\end{linenomath*}
where $M_i$ is the observed DT yield in the $i$th bin and $h_{\CP}$ is a normalization factor. DT yields with a $\CP$ eigenstate tag can only provide sensitivity to $c_i$.

The second type of tags are the multi-body decays $D \to \ks\pip\pim$ and $D \to \kl\pip\pim$, which provide sensitivity to $s_i$. The observed DT yields where the signal decays into bin $i$ and the tag decays into bin $j$ are given by 
\begin{widetext}
\begin{equation}
\label{eq:DTKspipi}
\begin{split}
M[S_i|(\ks\pi^+\pi^-)_j] = & h_{\ks\pi\pi}(T^S_{i}T^{\ks\pi\pi}_{-j}+T^S_{-i}T^{\ks\pi\pi}_{j}-2\sqrt{T^S_{i}T^{\ks\pi\pi}_{-j}T^S_{-i}T^{\ks\pi\pi}_{j}}(c^S_ic^{\ks\pi\pi}_j+s^S_is^{\ks\pi\pi}_j)), \\
M[S_i|(\kl\pi^+\pi^-)_j] = & h_{\kl\pi\pi}(T^S_{i}T^{\kl\pi\pi}_{-j}+T^S_{-i}T^{\kl\pi\pi}_{j}+2\sqrt{T^S_{i}T^{\kl\pi\pi}_{-j}T^S_{-i}T^{\kl\pi\pi}_{j}}(c^S_ic^{\kl\pi\pi}_j+s^S_is^{\kl\pi\pi}_j)),\\
\end{split}
\end{equation}
\end{widetext}
where $h_{\ks\pi\pi}$ and $h_{\kl\pi\pi}$ are normalization factors. 
For these tags, the hadronic parameters 
for both tag modes  
are taken from the results reported in Refs.~\cite{PhysRevD.82.112006,BESIII:2020khq,BESIII:2020hlg}.

The sample of DT events in which both $D$ mesons decay to the signal channel is also sensitive to the strong-phase parameters. However, with the current data set and the signal channel of $D \to \fpi$, the yields of this process are too small to be analyzed.

The $T^S$ parameters can be determined from data where the tag is a state of definite flavor which tags the signal decay as either a $\Dz$ or $\Dzbar$. 
The normalization factors in Eqs.~\eqref{eq:DTCP} and~\eqref{eq:DTKspipi} can be written as a combination of branching fractions, selection efficiencies and the integrated luminosity of the data set. In order to minimize systematic uncertainties, single-tag (ST) yields are used instead where one $D$ meson decays to a tag final state with no restrictions on the decay of the other $D$ meson. The use of ST yields removes the need to know the branching fraction, selection efficiencies and integrated luminosity. Further corrections to account for tag-dependent reconstruction effects and migration of events into different phase-space regions can be determined from the simulation. The determination of these normalization factors is described in Sec.~\ref{sec:crux}.

The choice of an appropriate partitioning of the phase space is important to avoid all strong-phase parameters with small values, thereby ensuring good sensitivity in the measurement of the angle $\gamma$.  

The phase space of $D\to\fpi$ can be described by five variables: \{$m_+,m_-,\trm{cos}\theta_+,\trm{cos}\theta_-,\phi$\}. Here, $m_+(m_-)$ is the invariant mass of the $\pip\pip(\pim\pim)$ pair,
$\theta_+(\theta_-)$ is the helicity angle of the $\pi^+\pi^+(\pi^-\pi^-)$ pair,
and $\phi$ is the angle between the $\pi^+\pi^+$ and $\pi^-\pi^-$ decay planes. The {\sc HyperPlot} package is used to create a model-implementation-independent scheme, referred to as a hyper-binning~\footnote{http://samharnew.github.io/HyperPlot/index.html}. To obtain square phase-space boundaries that can be easily used in the hyper-binning, 
the $m_\pm$ are transformed into $m'_\pm$ by
\begin{linenomath*}
\begin{equation}
m'_\pm = m_\pm + \delta, \delta = \trm{min}\{m_+,m_-\} - m_\trm{min},
\end{equation}
\end{linenomath*}
where $m_\trm{min}$ is the minimum kinematically possible value for $m_\pm$. With this choice of variables, the kinematically allowed region of \{$m'_+, m'_-, \trm{cos}\theta_+, \trm{cos}\theta_-, \phi$\} ranges from \{$m_\trm{min}, m_\trm{min}, -1, -1, -\pi$\} to \{$m_\trm{max}, m_\trm{max}, 1, 1, \pi$\}, where $m_\trm{max}$ is the maximum kinematically allowed value for $m_\pm$. However, there are symmetries of this system due to $\CP$-conjugation and identical-particle exchange which can be used to fold the phase space. The symmetries are
\begin{widetext}
    \begin{equation}
	\begin{split}
	\mathcal{CP}\{m'_+,m'_-,\trm{cos}\theta_+,\trm{cos}\theta_-,\phi\} & \rightarrow \{m'_-,m'_+,\trm{cos}\theta_-,\trm{cos}\theta_+,-\phi\}, \\
	\left[\pi^+_1 \leftrightarrow \pi^+_2 \right]\{m'_+,m'_-,\trm{cos}\theta_+,\trm{cos}\theta_-,\phi\} & \rightarrow \{m'_+,m'_-,-\trm{cos}\theta_+,\trm{cos}\theta_-,\phi-\pi\},  \\
	\left[\pi^-_1 \leftrightarrow \pi^-_2 \right]\{m'_+,m'_-,\trm{cos}\theta_+,\trm{cos}\theta_-,\phi\} & \rightarrow \{m'_+,m'_-,\trm{cos}\theta_+,-\trm{cos}\theta_-,\phi-\pi\}.
	\end{split}
	\end{equation}
  \end{widetext}
The identical-particle exchange symmetries allow a folding along the lines $\trm{cos}\theta_+ = 0$ and $\trm{cos}\theta_- = 0$ reducing the phase-space volume by a factor of four. 
The $\CP$ transformation with a changed bin number can further fold the phase space.
With this implementation, the hyper-binning scheme is reduced to the range \{$m_\trm{min}, m_\trm{min}, 0, 0, 0$\} to \{$m_\trm{max}, m_\trm{max}, 1,1,\pi$\}. The phase space is then divided into hyperpixels and groups of hyperpixels are combined to form the $i$th phase-space region. When generating the hyper-binning scheme with the {\sc HyperPlot} package, the minimum edge lengths are set as \{39~MeV$/c^2$, 39 MeV$/c^2$, 0.06, 0.06, 0.19 rad\}. These limits are chosen empirically, and balance the need for small edge lengths to minimize precision loss against the increased computation time required for a larger number of pixels. 
 
\begin{figure}[!htp]
\subfloat{\label{fig:CiSi_Model}\includegraphics[width=0.5\textwidth]{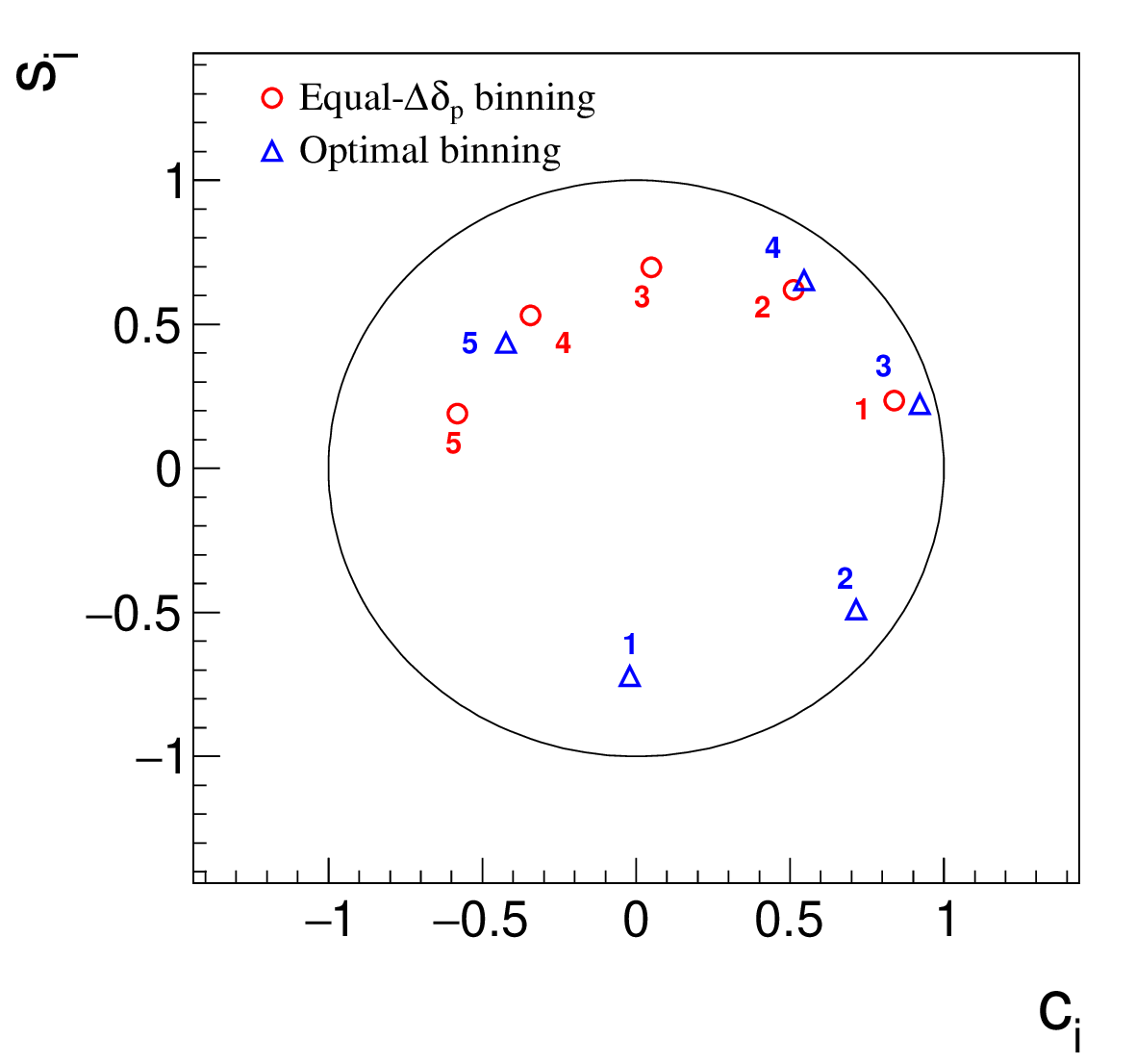}}
\hspace{0.05cm}
\subfloat{\label{fig:Ti_Model}\includegraphics[width=0.5\textwidth]{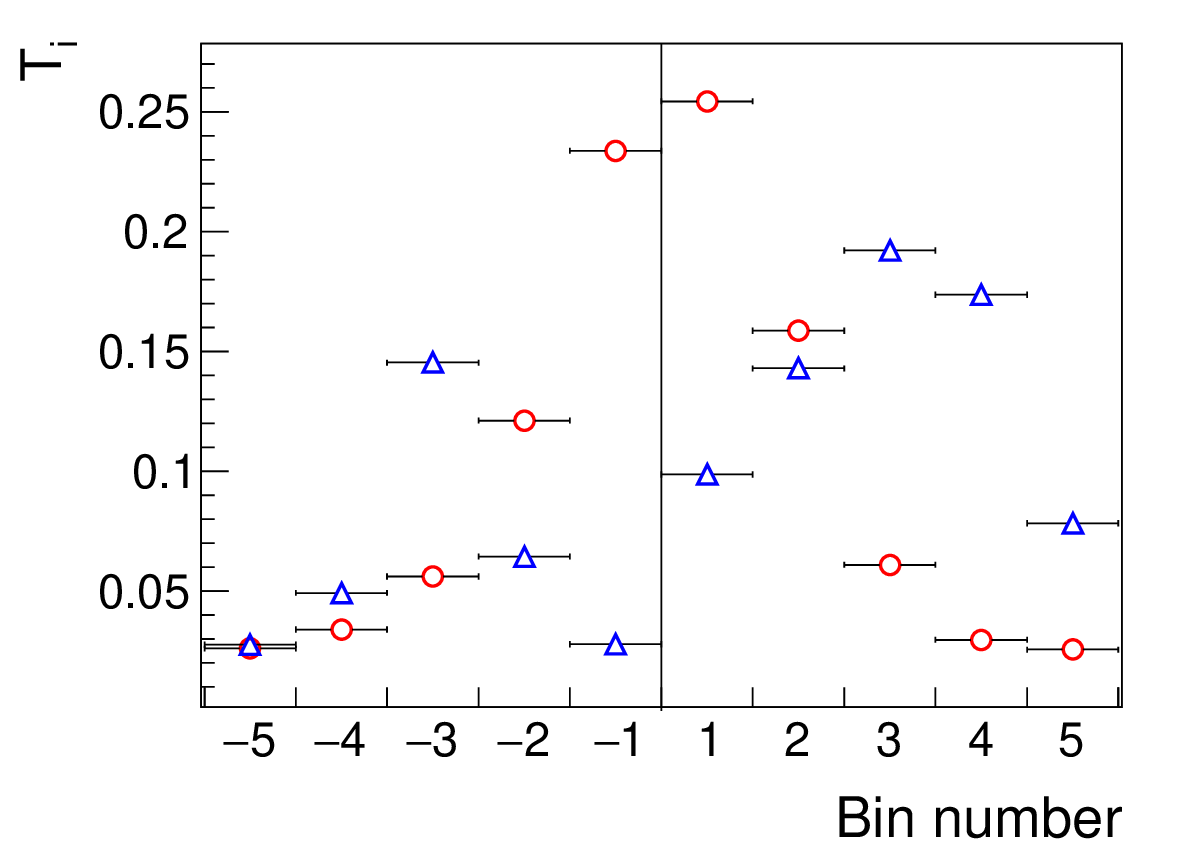}}
\caption[]{Model predictions of $c_i, s_i$ (top) and $T_i$ (bottom) for the two binning schemes.}
\label{fig:Model}
\end{figure}

Two schemes for the partitioning of phase space are presented~\cite{besiii:4pibins}. In each scheme, the phase-space region where the invariant mass of any $\pi^+\pi^-$ pair is in the range [0.481,0.514]~GeV$/c^2$ is removed to suppress the contamination from $D\to\ks\pi^+\pi^-$ decays. The remainder of the phase space is divided into 2$\times \mathcal{N}$ bins. In the first scheme, called the `equal-$\spp$ binning', hyperpixels are assigned to the $i$th bin based on the value of $\spp$ at the center of the pixel according to 
\begin{linenomath*}	
 \begin{equation}
	\label{eq:EqualBin}
	\begin{split}
	& +i : (i-1)\pi/\mathcal{N} \textless \spp  \textless i\pi/\mathcal{N}, \\
	& -i : -(i-1)\pi/\mathcal{N} \textgreater \spp \textgreater - i\pi/\mathcal{N},
 \end{split}
	\end{equation}
 \end{linenomath*}
 where $\spp$ is taken from the predictions of an amplitude model~\cite{BESIII:2023exz}. This choice groups together regions of similar strong phase to increase the coherence within each region and seeks to provide a simple bin definition that should lead to large values of $c_i$ and $s_i$. The value of $\mathcal{N}$ is set to five, primarily determined by the size of the data sample and how finely it can be partitioned for measurement. The predicted values of $c_i$ and $s_i$ are presented in Fig.~\ref{fig:Model} (top).
 The predicted points all lie close to the boundary of the unit circle, which is a consequence of the high level of coherence that exists in each bin, as predicted by the model.

The second scheme is referred to as the `optimal binning' since it is designed to provide the best statistical sensitivity in the $\gamma$  measurement. This choice takes into account the sensitivity of the rate equations for the decays used in the $\gamma$ measurement. The decay rates of the $B^\pm$ decays are written as
\begin{widetext}
\begin{equation}
\label{eq:BGamma}
\begin{split}
& \Gamma [B^-\to DK^-, D\to (\Fpi)_{\bf{p}}] \propto |\bar{A}^f_{\bf{p}}|^2r^2_B + |A^f_{\bf{p}}|^2 + 2|A^f_{\bf{p}}\bar{A}^f_{\bf{p}}|[x_-\trm{cos}(\spp)+y_-\trm{sin}(\spp)], \\
& \Gamma [B^+\to DK^+, D\to (\Fpi)_{\bf{p}}] \propto |A^f_{\bf{p}}|^2r^2_B + |\bar{A}^f_{\bf{p}}|^2 + 2|A^f_{\bf{p}}\bar{A}^f_{\bf{p}}|[x_+\trm{cos}(\spp)-y_+\trm{sin}(\spp)], \\
\end{split}
\end{equation}
\end{widetext}
where $x_\pm = r_B\trm{cos}(\delta_B\pm\gamma), y_\pm = r_B\trm{sin}(\delta_B\pm\gamma)$ and $r_B$ and $\delta_B$ are the amplitude ratio and strong-phase difference between the suppressed and favored $B$ decays.
Integrating the decay rate over the $i{\rm th}$ phase-space region gives the yields, $N$, observed in each such region for each $B$-meson charge,
\begin{linenomath*}
\begin{equation}
\label{eq:Byield}
\begin{split}
& N[B^-]_i \propto T_{-i}r^2_B + T_i + 2\sqrt{T_iT_{-i}}(c_ix_- + s_iy_-), \\
& N [B^+]_i \propto T_ir^2_B + T_{-i} + 2\sqrt{T_iT_{-i}}(c_ix_+ - s_iy_+). \\
\end{split}
\end{equation}
\end{linenomath*}

From these equations, it can be seen that the sensitivity to the interference term can be more dominant if $T_i$ and $T_{-i}$ are different in value and $r_B$ is small. Figure~\ref{fig:Model} (bottom) shows that in the equal-$\spp$-binning scheme, $T_i$ and $T_{-i}$ have similar values. In order to construct a scheme  with a large difference between $T_i$ and $T_{i}$, the hyperpixels are first divided such that 
\begin{linenomath*}
\begin{equation}
	\label{eq:AlterBin}
	\begin{split}
 & +i := \rfpi > 1 ,\\
 & -i := \rfpi < 1 ,
 \end{split}
\end{equation}
\end{linenomath*}
where $\rfpi = |A^{4\pi}_{\bf{p}}/\bar{A}^{4\pi}_{\bf{p}}|$. To account for the distribution of $D$-meson decays from the $B$-meson decay, a metric $Q_{\pm}$ is defined by
\begin{widetext}
\begin{equation}
	 Q^2_\pm = \frac{\sum_i \left(\frac{1}{\sqrt{N^i_{B^\pm}}} \frac{dN^i_{B^\pm}}{dx_\pm}\right)^2 + \left(\frac{1}{\sqrt{N^i_{B^\pm}}} \frac{dN^i_{B^\pm}}{dy_\pm}\right)^2}{\int_D \left[ \left( \frac{1}{\sqrt{\Gamma_{B^\pm}(\bf{p})}} \frac{d\Gamma_{B^\pm}(\bf{p})}{dx_\pm} \right)^2  +\left( \frac{1}{\sqrt{\Gamma_{B^\pm}(\bf{p})}} \frac{d\Gamma_{B^\pm}(\bf{p})}{dy_\pm} \right)^2   \right]{\rm d}\bf{p}}  .
   \label{eq:Qval}
	\end{equation}
 \end{widetext}
  Here, $N^i_{B^\pm}$ and $\Gamma_{B^\pm}(\bf{p})$ are defined in Eqs.~\eqref{eq:Byield} and~\eqref{eq:BGamma}, respectively, the values of $r_B$, $\delta_B$, and $\gamma$ are taken from Ref.~\cite{LHCbGammaCombo}, and $i$ is a Dalitz plot bin made up of hyperpixels.  This metric describes the expected statistical sensitivity of a measurement in relation to an unbinned analysis. 
  A higher value of $Q_{\pm}$ implies a higher sensitivity to $\gamma$.
  For the scheme to be useful experimentally, the overall phase space of each bin is required to be of sufficient size to have appreciable $B$-meson yields. Therefore, the final metric is:
  \begin{linenomath*}
\begin{equation}
	 Q'^2_{\pm} = Q^2_{\pm} - \frac{1}{10}\sum^\mathcal{N}_{i=1}
		 \begin{cases} 
		 T_i + \bar{T}_i < t : & \left[\frac{T_i+\bar{T}_i-t}{t}\right]^2  \\ 
	 	 T_i + \bar{T}_i > t : & 0 
		 \end{cases} , 
	\end{equation}
 \end{linenomath*}
where $t = \frac{2}{3\mathcal{N}}\sum^\mathcal{N}_{i=1}(T_i + \bar{T}_i)$ is the threshold which provides a penalty factor for small regions of phase space.

 The average value $Q'^2 = \frac{1}{2}(Q'^2_+ + Q'^2_-)$ is used to determine the best scheme. In practice, there are 280,000 hyperpixels and each hyperpixel can take one of ten possibilities, assuming that only five bin pairs are used. As it is not possible to compute all bin possibilities, an adaptive optimization procedure is used. The bin number of each hyperpixel is optimized one by one, cycling through all hyperpixels until the $Q'^2$ value no longer increases. 

The starting point of the algorithm is an equal-phase binning with the extra condition given in Eq.~(\ref{eq:AlterBin}). Inspection of the predicted values of $c_i$ and $s_i$ in Fig.~\ref{fig:Model} (top) shows that the bin values are now spread over the whole unit circle and closer to the boundary. The $Q'$ value for the equal-phase binning is 0.8 and $Q'=0.85$ for the optimal binning, which suggests that as long as the model is sufficiently accurate, these five-bin schemes will provide a minimal statistical sensitivity loss of (15-20)$\%$ on the measurement of $\gamma$ in comparison to an unbinned measurement. Although an unbinned approach based on an amplitude model has higher statistical precision, it introduces a model-dependent systematic uncertainty that is very difficult to quantify, making it unsuitable for a $\gamma$ measurement.

\section{Detector}
\label{detmc}

The BESIII detector~\cite{Ablikim:2009aa} records symmetric $e^+e^-$ collisions 
provided by the BEPCII storage ring~\cite{Yu:2016cof}, which operates with a peak luminosity of $1.1\times10^{33}$~cm$^{-2}$s$^{-1}$
in the center-of-mass (c.m.) energy range from 1.84 to 4.95~$\gev$.
BESIII has collected large data samples in this energy region~\cite{BESIII:2020nme}. The cylindrical core of the BESIII detector covers 93\% of the full solid angle and consists of a helium-based
 multilayer drift chamber~(MDC), a plastic scintillator time-of-flight
system~(TOF), and a CsI(Tl) electromagnetic calorimeter~(EMC),
which are all enclosed in a superconducting solenoidal magnet
providing a 1.0~T magnetic field. The solenoid is supported by an
octagonal flux-return yoke with resistive plate counter muon-identification modules interleaved with steel. 
The charged-particle momentum resolution at $1~{\rm GeV}/c$ is
$0.5\%$, and the 
${\rm d}E/{\rm d}x$
resolution is $6\%$ for electrons
from Bhabha scattering. The EMC measures photon energies with a
resolution of $2.5\%$ ($5\%$) at $1$~GeV in the barrel (end-cap)
region. The time resolution in the TOF barrel region is 68~ps, while
that in the end-cap region is 110~ps.

Simulated samples, which are produced with the {\sc geant4}-based~\cite{geant4} Monte Carlo (MC) package that includes the description of the detector geometry and response,
are used to determine the detection efficiencies and estimate the backgrounds.
The beam-energy spread of 0.97~MeV and the initial-state radiation (ISR) in the $\EE$ annihilations, which is modeled with the generator {\sc kkmc}~\cite{kkmc}, are included in the simulation.
The inclusive MC samples for background studies consist of the production of neutral and charged charm-meson pairs from $\psipp$ decays, decays of the $\psipp$ to charmonia or light hadrons, the ISR production of the $\jpsi$ and $\psip$ states and the continuum processes.  No attempt is made to include quantum-correlation effects in $\psipp \to D\bar{D}$ decays in the inclusive MC sample.
The equivalent integrated luminosity of the inclusive MC samples is about 10 times that of the data, except for the production of $D\bar{D}$ events, where the equivalent integrated luminosity is about 20 times that of the data.
All particle decays are modeled with {\sc
evtgen}~\cite{evtgen} using branching fractions either taken from the
Particle Data Group~\cite{pdg}, when available,
or otherwise estimated with {\sc lundcharm}~\cite{lundcharm,Yang:2014vra}.
The final-state radiation from the charged final-state particles is incorporated with the {\sc photos} package~\cite{photos}.

Signal MC samples of around 200,000 events are generated separately for the different tag channels. 
In this generation, the $D\to\fpi$ decay follows an isobar-based amplitude model fitted to a BESIII data sample of these decays~\cite{BESIII:2023exz}, where the flavor of the decaying meson is inferred by reconstructing the other charm meson in the event through its decay into a flavor-specific final state. The model contains the main resonant structures observed in the data.  The simulated DT samples involving $D \to \ksl\pipi$ tags are an order of magnitude larger in size and the tag decays are implemented with an amplitude model developed by the BaBar collaboration~\cite{BaBar:2010nhz}.  Quantum correlations are included in the generation of all the DT samples to ensure the best possible description of the reconstruction efficiency, especially for the different phase-space bins of the $D \to \ksl\pipi$ tag modes.

\section{Event selection}
\label{sec:selection}

Table~\ref{tab:TagInfo} lists the tag categories used in this analysis:
flavor-specific final states, $\CP$-even eigenstates, $\CP$-odd eigenstates, the quasi-$\CP$-even mode $D\to \pipi\piz$, and the self-conjugate decays $D\to \ksl\pipi$ of mixed $\CP$.  With the exception of the flavor-specific final states, all these tags were already used in the measurement of $F^{4\pi}_+$ reported in Ref.~\cite{BESIII:2022wqs}, where the selection and determination of the event yields are fully described.  Here, information is given on the selection of the additional tags and all event yields are summarized.  Charge conjugation is implicitly included throughout the discussion.

\begin{table}[!hbp]
\renewcommand\arraystretch{1.5}
\centering
\caption{Summary of tag channels.\vspace*{0.2cm}}
\label{tab:TagInfo}
\begin{tabular}{ l  c }
\toprule
Category & Decay mode\\
\hline
  Flavor 				 & $K^-\pi^+$, $K^-\pi^+\piz$, $K^-\pi^+\pi^-\pi^+$, $K^- e^+ \nu_e$  \\
	$\CP$ even    &  $K^{+}K^{-}$, $\ks\piz\piz$, $\kl\piz$, $\kl\omega$ \\
	$\CP$ odd     &  $\ks\piz$, $\ks\eta$, $\ks\eta'$, $\ks\omega$, $\kl\piz\piz$ \\
	Quasi-$\CP$ even & $\pipi\piz$  \\
	Mixed $\CP$ & $\ks\pipi$, $\kl\pipi$ \\
\bottomrule
\end{tabular}
\end{table}

The hadronic flavor tags are constructed from charged tracks and $\pi^0$ candidates that fulfill identical requirements to those in the previous analysis~\cite{BESIII:2022wqs}.   The positron used to form the $D \to K^- e^+ \nu_e$ candidates must pass the same charged-track quality criteria as the charged hadrons. The particle identification criteria for the positron are $P_e > P_K$ and $P_e > P_\pi,$ where  $P_x$ is the probability under the hypothesis $x$, constructed from measurements of the energy deposited in the MDC~(d$E$/d$x$) and the flight time in the TOF.

When selecting $D \to K^-\pi^+$ tags, it is demanded that the two charged tracks have a TOF time difference of less than 5~ns and that neither is identified as an electron or a muon.  These requirements suppress background from cosmic-ray and Bhabha events. To suppress background from $D \to K^\mp\ks\pi^\pm$ decays in the $D \to K^-\pip\pim\pip$ sample, a $\ks$ veto is applied, in which the event is rejected if the invariant mass of the $\pi^+\pi^-$ pair lies within the range [0.481, 0.514]~$\gevcc$.   As is the case for all the other fully reconstructed tags, a requirement is applied on the energy difference $\dE = E_D - \sqrt{s}/2$, to suppress the combinatorial background.  Here, $\sqrt{s}$ is the c.m. energy and $E_D$ is the measured energy of the $D$-meson candidate in the c.m. frame.   The value of $\dE$ is required to lie in the ranges $[-0.027, 0.025]$, $[-0.053,0.040]$, and $[-0.024,0.020]$~ GeV for the  $D \to K^-\pi^+$, $K^-\pi^+\piz$ and $K^-\pi^+\pi^-\pi^+$ candidates, respectively, which constitute windows of approximately three times the resolution of $\dE$ around zero.

\begin{figure}[!htp]
\begin{center}
\begin{overpic}[width=0.5\textwidth]{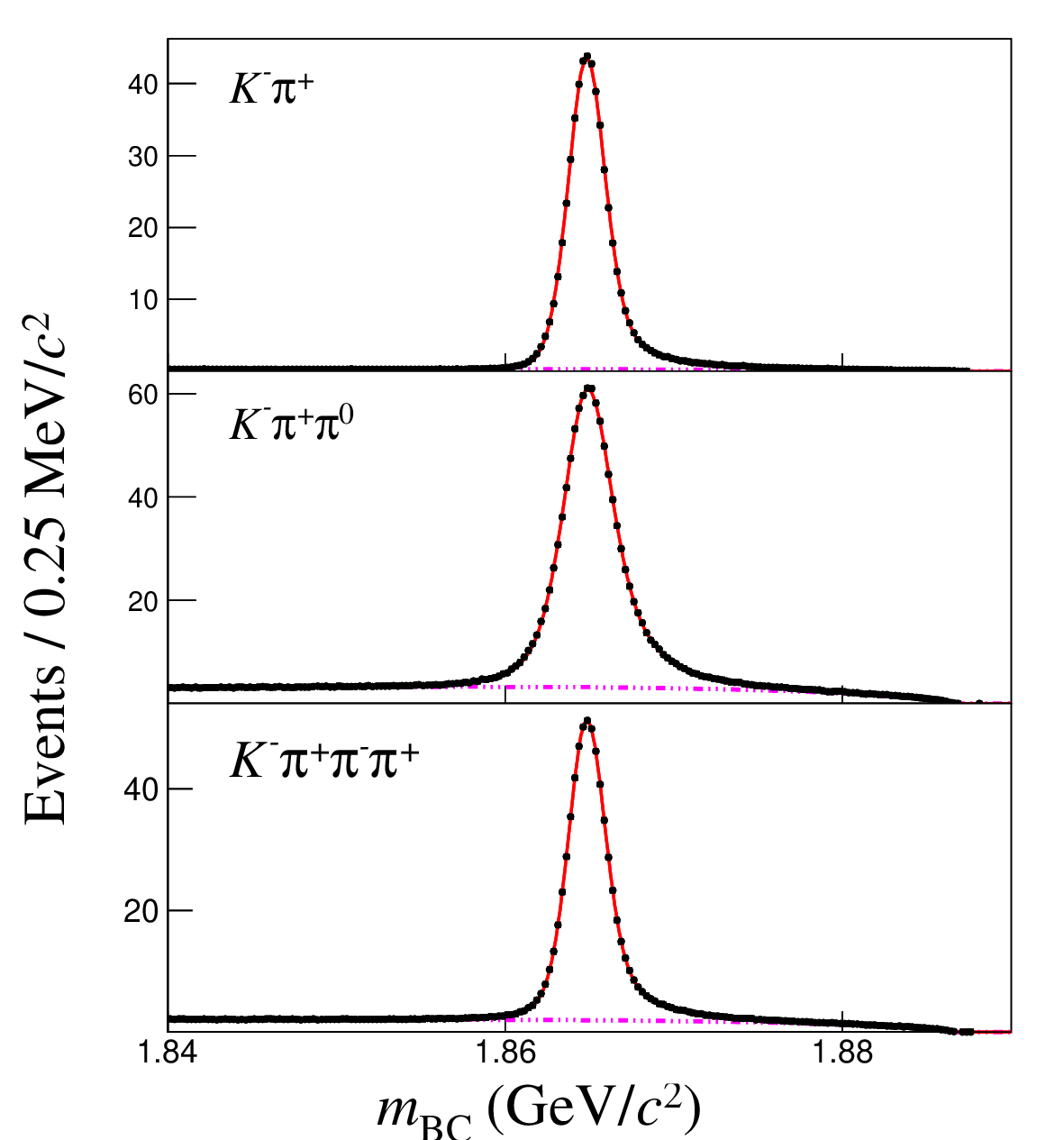}\put(16,98){{\sansmath $\times 10^3$}}\end{overpic}
\caption[]{Distributions and fits in $\mbc$ used to determine the ST yields for the hadronic flavor-tag channels.  The black dots with error bars are data, the total fit result is shown as the red solid line and the combinatorial background is shown as the pink dashed line. }
\label{fig:STFit}
\end{center}
\end{figure}

In order to determine the ST yields for the fully reconstructed channels, the beam-constrained mass, $\mbc = \sqrt{(\sqrt{s}/2)^2 - |\bf{p_D}|^2}$, is used to identify the signal, where $\bf{p_D}$ is the three-momentum vector of the $D$ candidate in the c.m. frame. 
Binned maximum-likelihood fits are performed on the $\mbc$ distributions, which are presented for the flavor tags in  Fig.~\ref{fig:STFit}. 
In the fit, the signal is described by the shape found in the MC simulation convolved with a Gaussian function to account for differences in resolution between MC and data, and the combinatorial background is described by an ARGUS function with an end-point fixed to $\sqrt{s}/2$~\cite{ARGUS:1990hfq}.
The peaking-background contributions from other charm decays are estimated from MC simulation and then subtracted from the fitted ST yields. 
For the flavor-tag channels, the fraction of the peaking background is less than 0.5\%.
The ST yields after background subtraction, $N_{\rm ST}$, are summarized in Table~\ref{tab:STDTYields}.

It is not possible to select a fully reconstructed ST sample for the mode $D \to K^- e^+ \nu_e$, nor are such samples available for decays involving a $\kl$ meson.  
However, effective ST yields can be determined through the relation
\begin{linenomath*}
\begin{equation}
\label{eq:KlST}
N_{\rm ST}(X) = 2N_{D\bar{D}} \, \BR (X) \, \epsilon_{\,\textrm{ST}}(X),
\end{equation}
\end{linenomath*}
where $N_{\rm ST}(X)$ is the effective ST yield for mode $D \to X$, $N_{D\bar{D}} = (10597\pm28\pm98)\times 10^3$ is the number of neutral charm-meson pairs in the sample~\cite{BESIII:2018iev} and
$\BR (X)$ is the branching fraction of the $D^0 \to X$ mode,  taken from Ref.~\cite{pdg} for $D \to K^- e^+ \nu_e$ and from Ref.~\cite{BESIII:2022qkh} for the modes involving a $K^0_L$ meson.
The effective efficiency $\epsilon_{\,\textrm{ST}}(X)$ is defined as the ratio of the efficiency for reconstructing $D \to \fpi$ versus $D \to X$ DT events with the partial-reconstruction technique, and the reconstruction efficiency for $D \to \fpi$.  The effective ST yields for these modes are included in Table~\ref{tab:STDTYields}~\footnote{The values for $\Nst$ for these tags were wrongly reported in Ref.~\cite{BESIII:2022qkh}, although the correct values, as given in Table~\ref{tab:STDTYields}, were used in the analysis.}.

\begin{table}[!hbp]
\renewcommand\arraystretch{1.5}
\centering
\caption{Summary of ST yields ($\Nst$), and DT yields ($\Ndt$) for each tag mode, grouped by tag category. The uncertainties are statistical, except for $\Nst$ in the case of tags involving a $\kl$, where they arise from the uncertainty in the contributing factors of Eq.~(\ref{eq:KlST}). The ST yields are not measured for $K^0_{S,L}\pi^+\pi^-$. 
 \vspace*{0.2cm}}
\footnotesize
\begin{tabular}{ l    r@{$\pm$}l r@{$\pm$}l}
\toprule
Mode  & \multicolumn{2}{c}{$\Nst$}  & \multicolumn{2}{c}{$\Ndt$}\\
\hline
$K^-\pi^+$ & 546675 & 775 &  1892.1&46.3 \\
$K^-\pi^+\piz$ & 1048420 & 1200 & 3288.9&63.0 \\
$K^-\pi^+\pipi$ & 655533 & 927 & 1895.3&49.6 \\ 
$K^- e^+ \nu_e$ & 424140 & 5850 & 1559.7 & 42.6\\ 
\multicolumn{5}{c}{} \\
$K^{+}K^{-}$ & 56668&262 & 115.4&14.4\\
$\ks\piz\piz$& 73176&299 & 36.4&10.3\\
$\kl\piz$    & 74830&2320& 130.9&18.8\\
$\kl\omega$  & 33210&1830& 61.5&13.8\\
\multicolumn{5}{c}{} \\
$\ks\piz$		 & 73176&299 & 326.0&19.2\\
$\ks\eta_{\gamma\gamma}$ & 10071&123 & 57.7&7.7\\
$\ks\eta_{\pip\pim\piz}$ & 2775&65 & 16.5&4.2\\
$\ks\eta'_{\pip\pim\eta}$ & 3449&67 & 11.6&3.5\\
$\ks\eta'_{\gamma\pip\pim}$ & 8691&126 & 41.1&7.5\\
$\ks\omega$  & 26220&215 & 128.7&13.8\\
$\kl\piz\piz$ & 30980&1230& 178.5&23.9\\
\multicolumn{5}{c}{} \\
$\pipi\piz$ & 115556&682 & 190.7&24.6\\
\multicolumn{5}{c}{} \\
$\ks\pipi$ &   \multicolumn{2}{c}{/}   & 539.7&26.0\\
$\kl\pipi$ &   \multicolumn{2}{c}{/}   & 1374.6&50.4\\
\bottomrule
\end{tabular}
\label{tab:STDTYields}
\end{table}

\begin{figure*}[!htp]
\begin{center}
\begin{overpic}[width=1.0\textwidth]{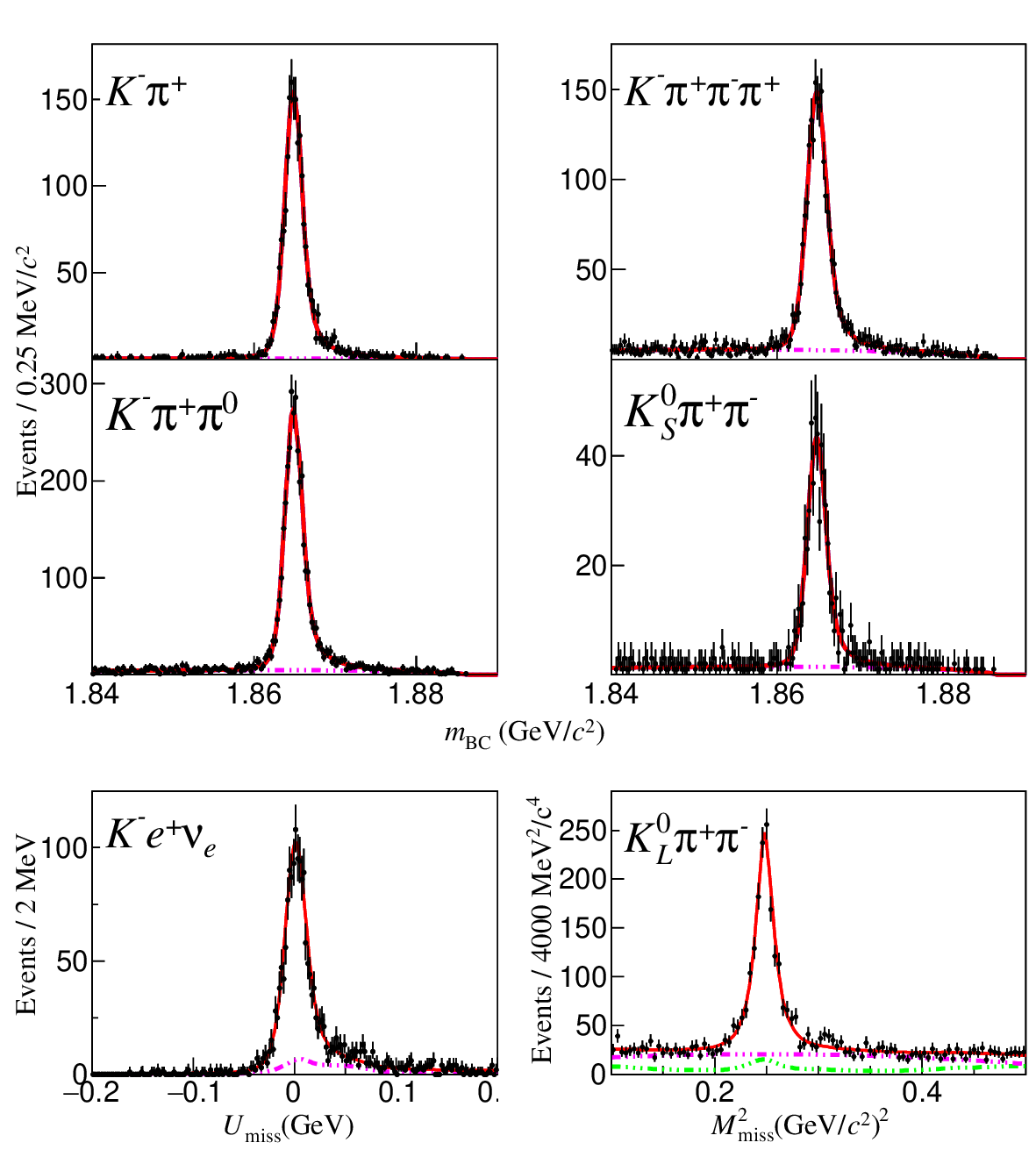}\end{overpic}
\caption[]{Distributions and fits used to determine the DT yields for four fully reconstructed tags, and for the partially reconstructed tags  $K^- e^+ \nu_e$ and $\kl\pipi$.  These distributions are of $\mbc$ in the fully reconstructed case, and of $\umiss~(\missm)$ for the $K^- e^+ \nu_e~(\kl\pipi)$ case. 
In each plot the black dots with error bars are data, 
the total fit result is shown as the red solid line,
the combinatorial background is shown as the pink dot-dashed line, and the sum of the combinatorial and the peaking background for the $D \to K^-e^+\nu_e$ and $\kl \to \pipi$ DTs is shown as the green dashed line.
}
\label{fig:DTFit}
\end{center}
\end{figure*}

Samples of DT events are selected for all tags not involving a $\kl$ or $\nu_e$ by attempting to reconstruct the $D \to \fpi$ signal decay from the remaining tracks in the ST samples, as described in Ref.~\cite{BESIII:2022wqs}.
The $\ks$ veto, $[0.481, 0.514]~\gevcc$, is applied to suppress background from $D\to\ks\pipi$ decays, as was done in Ref.~\cite{BESIII:2022wqs}.
The DT yield is determined by an unbinned maximum-likelihood fit to the $\mbc$ distribution of the signal $D$ candidate. 
The $\mbc$ distributions and the fits for the fully reconstructed flavor modes and $D \to \ks\pipi$ tags are presented in Fig.~\ref{fig:DTFit}.  In these fits, the signal is described by the MC-simulated shape convolved with a Gaussian function, and the combinatorial background is described by an ARGUS function.
The fitted signal yield includes contamination from peaking-background contributions, which is dominated by $\sim 5\%$ residual contamination from $D \to \ks\pipi$ decays in the selection of the signal channel. 
The DT yields after background subtraction are summarized in Table~\ref{tab:STDTYields}.

The DT channels involving a $D \to K^- e^+ \nu_e$ or $D \to \kl X$  tag mode cannot be fully reconstructed.  Nonetheless, a partial reconstruction is performed, accounting for all other charged and neutral particles in the event, and the yield is determined using a fit to the missing energy or missing invariant-mass squared in the event.  In this procedure, the $D\to\fpi$ candidate is first reconstructed with the same criteria as used previously, and then the standard selections are imposed to select the remaining charged and neutral tracks in the event. 
$D \to K^- e^+ \nu_e$ candidates with any additional charged tracks or good $\piz$, apart from the kaon and positron candidates, are discarded.
 The signal yields are determined by performing unbinned maximum-likelihood fits to the variable $\umiss = \sqrt{s}/2-E_X-|{\bf p_{X}}+{\bf \hat{p}_{\Fpi}}\sqrt{s/4-M^2_D}|$ for the $D \to K^- e^+ \nu_e$ tag mode and the squared missing-mass 
 $\missm = (\sqrt{s}/2-E_X)^2 - |{\bf p_{X}}+{\bf \hat{p}_{\Fpi}}\sqrt{s/4-M^2_D}|^2$ for the $D \to \kl X$ tag modes, both calculated in the c.m. frame. Here, $E_X$ and ${\bf p_{X}}$ are the energy and three-momentum of the reconstructed particles in the event not associated with the signal candidate, ${\bf \hat{p}_{\Fpi}}$ is the unit vector in direction of the signal candidate momentum, and $M_D$ is the known $\Dz$ mass~\cite{pdg}.
 The $\umiss$ and $\missm$ distributions and accompanying fits are shown in  Fig.~\ref{fig:DTFit} for the $D \to K^- e^+ \nu_e$ and $D \to \kl \pipi$ tags. In these fits, the signal distribution is modeled by the shape obtained from the MC simulation convolved with a Gaussian function, where the width is a free parameter. The shape and size of contributions from the peaking background are fixed according to the MC simulations, and the combinatorial background is modeled with a first-order polynomial whose parameters are determined in the fit. Peaking backgrounds mainly arise from $D\to\ks\pipi$ decays misreconstructed as $D \to \pipi\pipi$ or $D\to\ks(\piz\piz)\pipi$ in the selection of the $D \to \kl\pipi$ channel. 
 The resulting signal yields are listed in Table~\ref{tab:STDTYields}.

\section{Determination of the hadronic parameters}
\label{sec:meas}

To determine the flavor-tag fractions and strong-phase parameters, the yield fits are performed in the phase-space bins, which are defined according to the equal-$\spp$-binning and optimal-binning schemes introduced in Sec.~\ref{sec:formalism}.  The four-momenta of the final-state particles of the $D \to \fpi$ decay, which are used to define the phase-space bin, are taken from the output of a kinematic fit that imposes the $D$-meson mass as a constraint. The same procedure is followed for the $D \to K^0_{S}\pi^+\pi^-$ tags, which are analyzed in their own phase-space bins. 

The fit to determine the $c_i$ and $s_i$ parameters is performed using a Poisson likelihood fit where the observed signal yield in each DT bin is compared to the expected value. 

\subsection{Measurement of the flavor-yield fractions}

The observed DT yields in bin $i$ of phase space are determined separately for each flavor-tag mode: $K^-\pi^+$, $K^-\pi^+\piz$, $K^-\pi^+\pi^-\pi^+$, and $K^- e^+\nu_e$. Peaking backgrounds are subtracted based on their measured branching fractions~\cite{pdg}.
The dominant background contribution to the signal sample is from $D \to \ks\pipi$ decays. The distribution of this background is modeled using the amplitude model described in Ref.~\cite{BaBar:2010nhz}.

For the three hadronic flavor-tagged DT samples, the decays are driven not only by the Cabibbo-favored (CF) amplitude but also by the doubly Cabibbo-suppressed (DCS) amplitude and their interference.
The effect is accounted for by multiplying the background-subtracted DT yield by the correction factor
\begin{linenomath*}
\begin{equation}
\label{eq:flavorCorrection}
f_i = \frac{\int_i |A_{\bf p}|^2\mathrm{d}\textbf{p}}{\int_i (|A_{\bf p}|^2 + (r^F_D)^2|\bar{A}_{\bf p}|^2 - 2r^F_D R_F\mathcal{R}[e^{i\delta_D^F}A_{\bf p}\bar{A}_{\bf p}])\mathrm{d}\textbf{p}},
\end{equation}
\end{linenomath*}
where the integrals are over the phase space of the signal decay in bin $i$, with the amplitude $A_p$ taken from the model of Ref.~\cite{BESIII:2023exz}. The parameter $r^F_D$ is the ratio of the magnitude of the DCS amplitude to that of the CF amplitude, and $\delta_D^F$ is the strong-phase difference between the two amplitudes.
In the case of the $D \to K^-\pi^+\pi^0$ and $D \to K^- \pi^+\pi^-\pi^+$ tags, $r^F_D$ and $\delta_D^F$ are averaged over the phase space of the decays, and the coherence factor $R_F$ accounts for the dilution in the interference caused by the intermediate resonances of these decays.  For the two-body decay $D \to K^- \pi^+$, the coherence factor is unity. The values of $r_D^F$, $\delta_D^F$ and $R_F$ are derived from measurements of quantum-correlated $D\bar{D}$ decays and charm mixing, which are summarized in Table~\ref{tab:ExternalInputs}.

\begin{table}[!hbp]
\renewcommand\arraystretch{1.5}
\centering
\caption{Values of the input parameters used to calculate the DCS correction factors for the hadronic flavor-tag modes.  The source of each set of numbers is listed under `Ref.'. }
\label{tab:ExternalInputs}
\begin{tabular}{ c c c  c  c}
	\toprule
	Tag mode  & Ref.  &  $r^F_D(\%)$ &  $\delta_D^F(^\circ)$ & $R_F$\\
	\hline
	$K^-\pi^+$ & \cite{HeavyFlavorAveragingGroup:2022wzx} & 5.86$\pm$0.02 & 191.7$\pm 3.7$ & 1 \\
	$K^-\pip\piz$ & \cite{BESIII:2021eud} & 4.41$\pm$0.11  & 196$\pm 11$ & 0.79$\pm$0.04 \\
	$K^-\pip\pim\pip$ & \cite{BESIII:2021eud} & 5.50$\pm0.07$  & 161$^{+28}_{-18}$ & 0.44$^{+0.10}_{-0.09}$ \\
	\bottomrule
\end{tabular}
\end{table}

The measurement of the position of the signal decay in phase space is imperfect due to the finite detector resolution, and hence some decays are assigned to the wrong bin in phase space. This migration from true to reconstructed bin is accommodated through an efficiency matrix, determined from MC simulation, with elements: 
\begin{linenomath*}
\begin{equation}
\label{eq:EffMatrix}
\epsilon_{ij} = \frac{N^\textrm{rec}_{ij}}{N^\textrm{gen}_{j}\epsilon_{\textrm{ST}}},
\end{equation}
\end{linenomath*}
where $N^\textrm{gen}_{j}$ is the number of MC events generated in the $j$th bin, $N^\textrm{rec}_{ij}$ is the number of MC events generated in the $j$th bin but reconstructed in the $i$th bin, and $\epsilon_{\textrm{ST}}$ is the ST efficiency of the flavor-tag channel.  
The diagonal elements vary between 31\% and 43\%,  while the off-diagonal elements are mostly less than 1\% and always less than 3\% for the four DT selections.

In order to handle the normalization between the observed and expected yields, it is useful to define the $K^S_i$ parameters. These are related to the $T_i$ via $T_i = K^S_i/\sum_{i=1}^{5}(K^S_i + \bar{K^S}_i)$. Inverting the efficiency matrix allows the $K_i^S$ to be determined through 
\begin{linenomath*}
\begin{equation}
K^S_i = \sum_{\mathrm{all\ bins}} (\epsilon^{-1})_{ji} (N_j^{\textrm{obs}}),
\end{equation}
\end{linenomath*}
where the sum runs over the five negative and five positive bins, and $N_j^{\textrm{obs}}$ is the observed DT yield corrected for all effects discussed above.  

The numerical results for the average over all tags for $T_i$ and $K^S_i$ are given in Table~\ref{tab:Ti} in the appendix. 
 Figure~\ref{fig:Ti} shows the results for $T_i$ for each tag, for both binning schemes.  The results for each tag are compatible.  

\begin{table*}[h]
\renewcommand\arraystretch{1.5}
    \centering
    \caption{Measured values of $T_i$ and $K^S_i$ averaged over all flavor tags.}
    \label{tab:Ti}
    \begin{tabular}{ c  r@{$\pm$}l  r@{$\pm$}l  r@{$\pm$}l  r@{$\pm$}l }
          \toprule                    
& \multicolumn{4}{c}{Equal-$\sp$-binning scheme} & \multicolumn{4}{c}{Optimal-binning scheme}\\  
  \hline 
$i$ & \multicolumn{2}{c}{$T_i$} & \multicolumn{2}{c}{$K^S_i$} & \multicolumn{2}{c}{$T_i$} & \multicolumn{2}{c}{$K^S_i$} \\
  \hline 
$-$5 & 0.023&0.002 & 464.0&45.6 & 0.021&0.003 & 413.4&50.9\\  
$-$4 & 0.031&0.003 & 619.4&51.9 & 0.041&0.003 & 823.3&59.4 \\  
$-$3 & 0.049&0.003 & 967.8&64.8 & 0.150&0.006 & 2973.6&97.5 \\  
$-$2 & 0.127&0.005 & 2505.7&93.2 & 0.056&0.004& 1119.9&65.3  \\  
$-$1 & 0.242&0.006 & 4780.7&121.4 & 0.021&0.002& 420.6&45.1 \\  
$\phantom{-}$1 &  0.256&0.007 & 5054.4&126.1 &0.099&0.004 & 1964.6&81.4 \\  
$\phantom{-}$2 &  0.156&0.005 & 3086.4&102.6 & 0.152&0.005& 3015.9&98.0\\  
$\phantom{-}$3 &  0.059&0.003 & 1174.5&64.8 & 0.194&0.006 & 3865.9&110.0 \\  
$\phantom{-}$4 &  0.031&0.003 & 607.2&52.1 & 0.184&0.005 & 3662.2&107.3\\  
$\phantom{-}$5 &  0.026&0.002 & 516.8&46.4 & 0.082&0.004 & 1627.7&73.7\\  
  \bottomrule 
    \end{tabular}
\end{table*}

\begin{figure}[]
\subfloat{\label{fig:Ti_Equal}\includegraphics[width=0.5\textwidth]{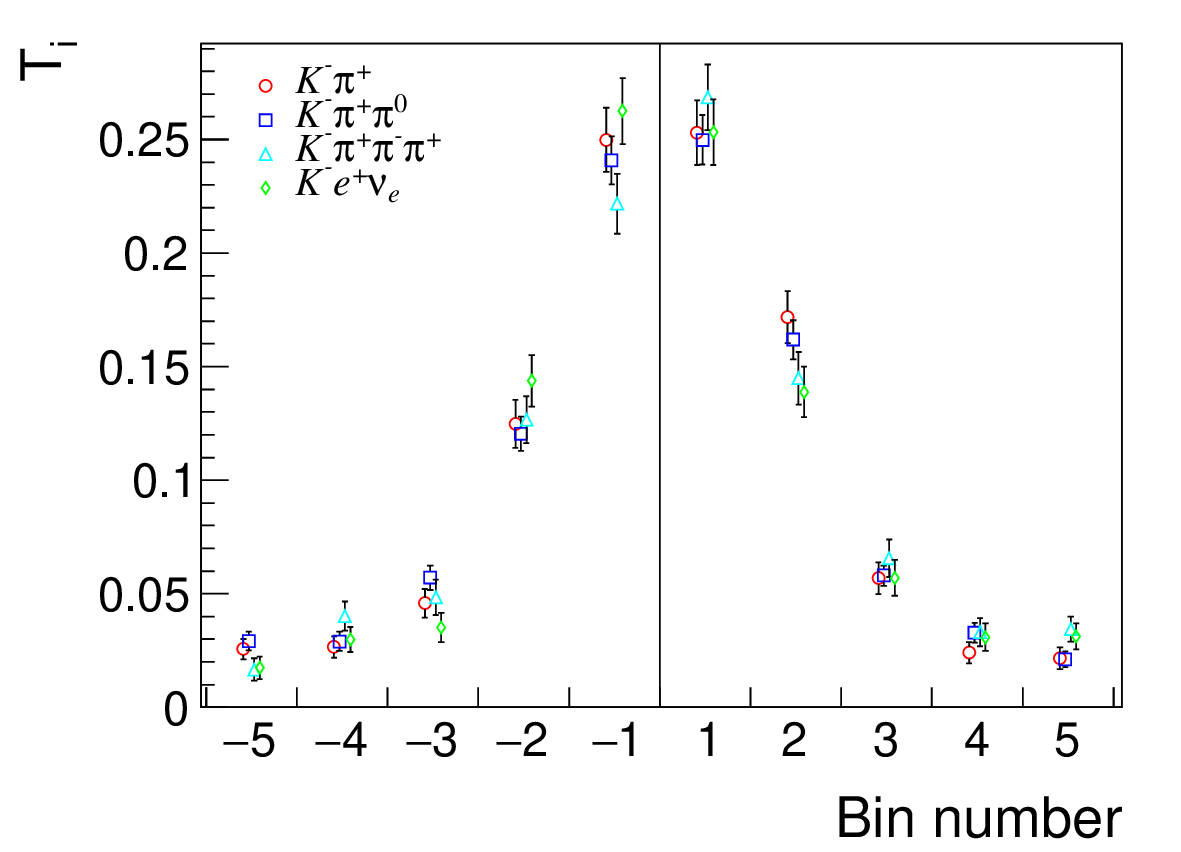}}   
\hspace{0.05cm}
\subfloat{\label{fig:Ti_Optim}\includegraphics[width=0.5\textwidth]{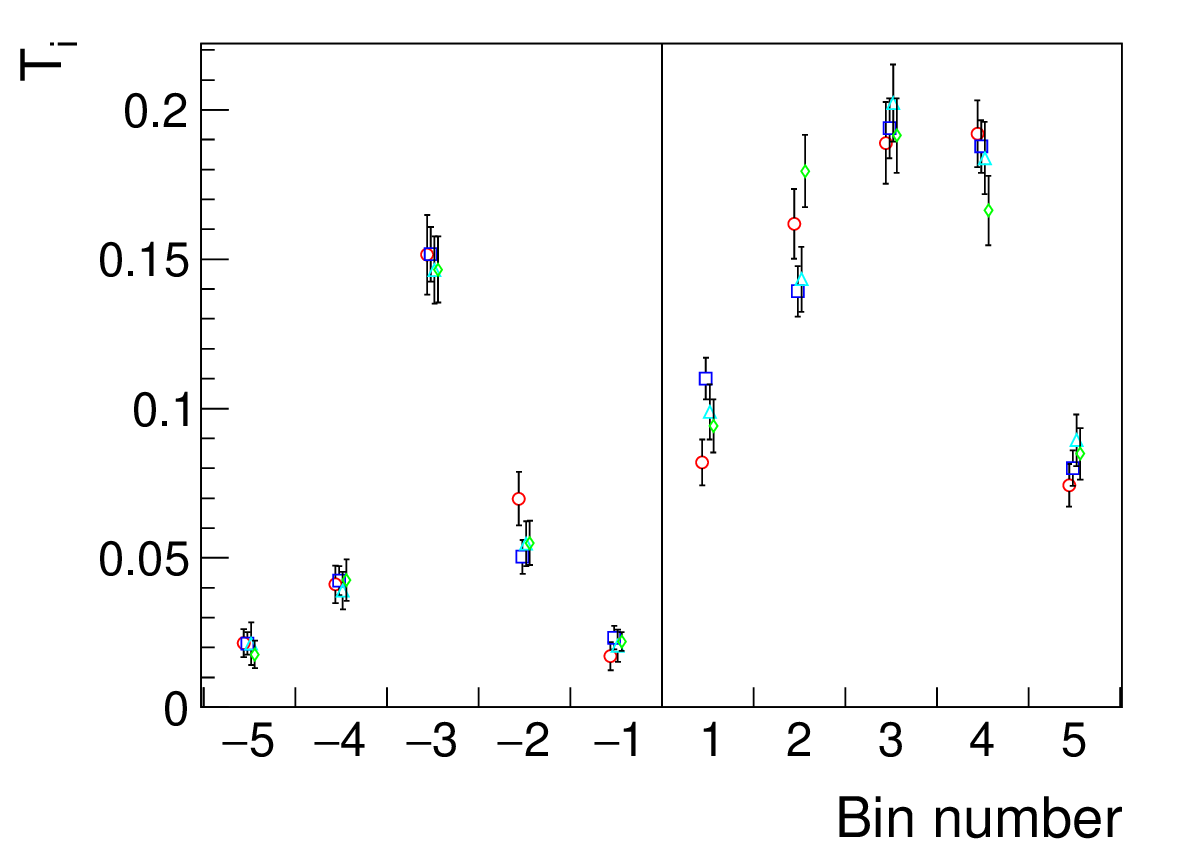}}  
\caption[]{Comparison of $T_i$ between different flavor-tag modes. The top plot shows the equal-$\spp$ binning and the bottom one the optimal binning.}
\label{fig:Ti}
\end{figure}

\subsection{Measurement of the strong-phase parameters $c_i$ and $s_i$}
\label{sec:crux}

The equations in Sec.~\ref{sec:formalism} require modification to take into account the experimental effects of selection efficiency and bin migration. The observed DT yields in bins across the different tags are compared to the expected yield, and the $c_i$ and $s_i$ parameters are determined through the minimization of the Poisson likelihood. Symmetries are exploited to merge bins. In the $\CP$ and quasi-$\CP$ tags, the expected yields are the same under the exchange $i\leftrightarrow -i$, as is evident from Eq.~(\ref{eq:DTCP}). Therefore, the yields in these pairs of bins are aggregated. The expected observed signal yield in bin $i$ against a $\CP$ tag, $N^{S|\CP}_i$ is given by
\begin{widetext}
\begin{equation}
    N^{S|\CP}_i= \left(\sum_{j=1}^{5}\epsilon_{ij} \frac{N_{\rm ST}^{\CP}}{2N_{\rm ST}^{\rm FT}(1-\eta_{\CP}y)}\left[K^S_j + K^S_{-j} -2\eta_{\CP}c_j \sqrt{K^S_jK^S_{-j}}\right]\right) + N_i^{\mathrm{bkg}},
    \label{eq:cptagwcorr}
\end{equation}
\end{widetext}
where $\epsilon_{ij}$ takes into account the selection efficiency and bin migration in the same way as in Eq.~(\ref{eq:EffMatrix}) and is determined by simulation, separately for each $\CP$ tag to accommodate any tag-dependent effects in the reconstruction efficiency. The parameter $N_{\rm ST}^{\CP}$ is the ST yield of the $\CP$ tag, and $N_{\rm ST}^{\rm FT}$ is the sum of the ST yields for the four flavor tags, which is 2674602$\pm$ 6093. The factor $(1-\eta_{\CP} y)$, where $\eta_{\CP}$\ is the eigenvalue of the tag and $y=0.00630$ is the charm-mixing parameter~\cite{HeavyFlavorAveragingGroup:2022wzx}, takes into account a correction that relates the ST yield to the branching fraction in the quantum-correlated data. The fitted yields include a peaking background and hence the expected size of this contribution is included as a term in Eq.~(\ref{eq:cptagwcorr}). As discussed in Ref.~\cite{BESIII:2022qkh}, the peaking background is mainly from $D\to\ks\pipi$ decays in the selection of the signal channel except for the $D \to \pipi\piz$ tag channel where there are $D\to\ks\piz$ events in the selected $D \to \pipi\piz$ events. 
The peaking-background fraction in each bin is estimated through simulation in which the $D\bar{D}$ pair is quantum-correlated. Equation~\eqref{eq:cptagwcorr} is modified for the $D \to \pip\pim\pi^0$ tag through the replacement of $\eta_{\CP}\leftrightarrow (2F_+^{\pi\pi\pi^0}-1)$.

In the case of the $D\to\fpi$ versus $D\to K^0_{S,L}\pipi$ DTs, both the signal and tag channels are divided into bins of phase space. The binning of the tag decay is the same for both $D\to\ks\pipi$ and $D\to\kl\pipi$, following the `equal-phase' scheme as defined in Ref.~\cite{BESIII:2020khq} and consists of 2$\times 8$ bins. The strong-phase parameters $c^{\ks\pi\pi}_i$, $s^{\ks\pi\pi}_i$, $c^{\kl\pi\pi}_i$ and $s^{\kl\pi\pi}_i$ for $D\to K^0_{S,L}\pipi$ are taken from Ref.~\cite{BESIII:2020khq} and are the combined results from the BESIII and CLEO measurements~\cite{BESIII:2020khq,PhysRevD.82.112006}. The flavor-tag yields $K^{\ks\pi\pi}_i$ and $K^{\kl\pi\pi}_i$ are also taken from Ref.~\cite{BESIII:2020khq}, determined using only BESIII data. When the signal decays to bin $i$ and the tag to bin $j$, the expected yield is symmetric under the exchange of  $(i,j)\leftrightarrow (-i,-j)$. Therefore, the observed yields are combined, resulting in 80 distinct bin combinations for each tag. For each tag, an 80$\times$80 efficiency and migration matrix is constructed from simulation to account for the true $(i,j)$ bin being reconstructed in the $(l,m)$ bin combination. The peaking background expected in each bin pair, $N_{ij}^{\mathrm{bkg}}$, is estimated by a simulation including quantum correlations. The expected observed yield is then given by
\begin{linenomath*}
\begin{widetext}
    \begin{equation}
        N_{i,j}^{S|T}=\left(\sum_{l=-5}^{l=+5} \sum_{m=-8}^{m=8} \epsilon_{ijlm}\frac{N_{D\bar{D}}}{N_{\rm ST}^{\rm FT}N_{\rm ST}^{{\rm FT},T}}\left[K^S_lK^T_{-m}+K^S_{-l}K^T_{m} - 2\alpha\sqrt{K^S_lK^T_{-m}K^S_{-l}K^T_{m}}\left(c_lc^T_m+s_ls^T_m\right)\right]\right) + N_{ij}^{\mathrm{bkg}},
        \label{eq:k0pipiwcor}
    \end{equation}
\end{widetext}
\end{linenomath*}
where $T$ designates the tag mode and the factor $\alpha$ =1 for the $D \to \ks\pipi$ tag and $\alpha=-1$ for the $D \to \kl\pipi$ tag. The normalization factor includes $N_{\rm ST}^{{\rm FT},T}$ which is the ST yield for the sum of flavor tags used to determine the $K^T_i$. These yields are reported in Ref.~\cite{BESIII:2020khq} as $2796832\pm 6182$ for the $D\to\ks\pipi$ tag and $2337843\pm 2335$ for the $D\to\kl\pipi$ tag.

The Poisson probability of observing $N$ events given an expectation value of $\langle N\rangle$ is
\begin{linenomath*}
\begin{equation}
P(N;\langle N\rangle) = \frac{\langle N\rangle^N e^{-\langle N\rangle}}{N!}.
\end{equation}
\end{linenomath*}
The $c_i$ and $s_i$ are obtained by minimizing the negative log-likelihood expression
\begin{linenomath*}
\begin{equation}
\label{eq:CiSiLog}
\begin{split}
%
-2\rm{log}\mathcal{L} = & -2\mathop{\sum}_{\CP~\rm{tag}}\mathop{\sum}_{i} \textrm{log}~P\left(N_i^{\mathrm{obs}:S|\CP};\langle N_i^
{S|\CP}\rangle\right)  \\
& -2\mathop{\sum}_{i} \textrm{log}~P\left( N_i^{\mathrm{obs}:S|\pi\pi\piz};\langle N_i^{S|\pi\pi\piz}\rangle\right) , \\
& -2\mathop{\sum}_{ij} \textrm{log}~P\left(N_{ij}^{\mathrm{obs}:S|\ks\pi\pi};\langle N_{ij}^{S|\ks\pi\pi}\rangle\right) ,\\
& -2\mathop{\sum}_{ij}\textrm{log}~P\left(N_{ij}^{\mathrm{obs}:S|\kl\pi\pi};\langle N_{ij}^{S|\kl\pi\pi}\rangle\right) ,
\end{split}
\end{equation}
\end{linenomath*}
\noindent where $\langle N\rangle$ is the expected yield, and $N^{\mathrm{obs}}$ is the observed yield in the data sample. 

Samples of simulated pseudodata are used to validate the fit procedure and no bias is found.
The fit is performed using the two binning schemes and the results for the $c_i$ and $s_i$ are summarized in Table~\ref{tab:CiSiResults} as listed in the appendix. 
The corresponding correlation matrices are provided in the Appendix. The results are also displayed in Fig.~\ref{fig:CiSi}. Although the model is not confirmed by the measurements in all bins, the general pattern of agreement gives confidence that each binning scheme provides good sensitivity in the $\gamma$ measurement. 

\begin{table*}[htbp]
\renewcommand\arraystretch{1.5}
  \centering
  \caption{Summary of $c_i$ and $s_i$ results, where the first uncertainties are statistical and the second uncertainties are systematic.}
  \begin{tabular}{l c c  c c }
  \toprule
    & \multicolumn{2}{c}{Equal-$\sp$-binning scheme} & \multicolumn{2}{c}{Optimal-binning scheme} \\
    $i$ & $c_i$ & $s_i$ &$c_i$ & $s_i$\\
  \hline 
  1 & $\phantom{-}$0.798$\pm$0.022$\pm$0.008 & $-$0.116$\pm$0.108$\pm$0.029  & $\phantom{-}$0.119$\pm$0.091$\pm$0.021 & $-$0.424$\pm$0.210$\pm$0.043  \\
  2 & $\phantom{-}$0.406$\pm$0.041$\pm$0.010 & $\phantom{-}$0.546$\pm$0.137$\pm$0.036 & $\phantom{-}$0.738$\pm$0.044$\pm$0.017 &$-$0.390$\pm$0.161$\pm$0.058  \\
  3 & $\phantom{-}$0.262$\pm$0.077$\pm$0.017 & $\phantom{-}$0.777$\pm$0.187$\pm$0.050 & $\phantom{-}$0.808$\pm$0.027$\pm$0.012 & $-$0.250$\pm$0.124$\pm$0.030  \\
  4 & $-$0.301$\pm$0.097$\pm$0.031 & $-$0.669$\pm$0.343$\pm$0.095 & $\phantom{-}$0.423$\pm$0.059$\pm$0.017 & $\phantom{-}$0.857$\pm$0.186$\pm$0.074 \\
  5 & $-$0.585$\pm$0.109$\pm$0.029 & $\phantom{-}$0.225$\pm$0.321$\pm$0.092 & $-$0.273$\pm$0.094$\pm$0.025 & $-$0.225$\pm$0.252$\pm$0.081  \\
  \bottomrule
  \end{tabular}
  \label{tab:CiSiResults}
\end{table*}

\begin{figure}[!htp]
\subfloat{\label{fig:CiSi_Equal}\includegraphics[width=0.5\textwidth]{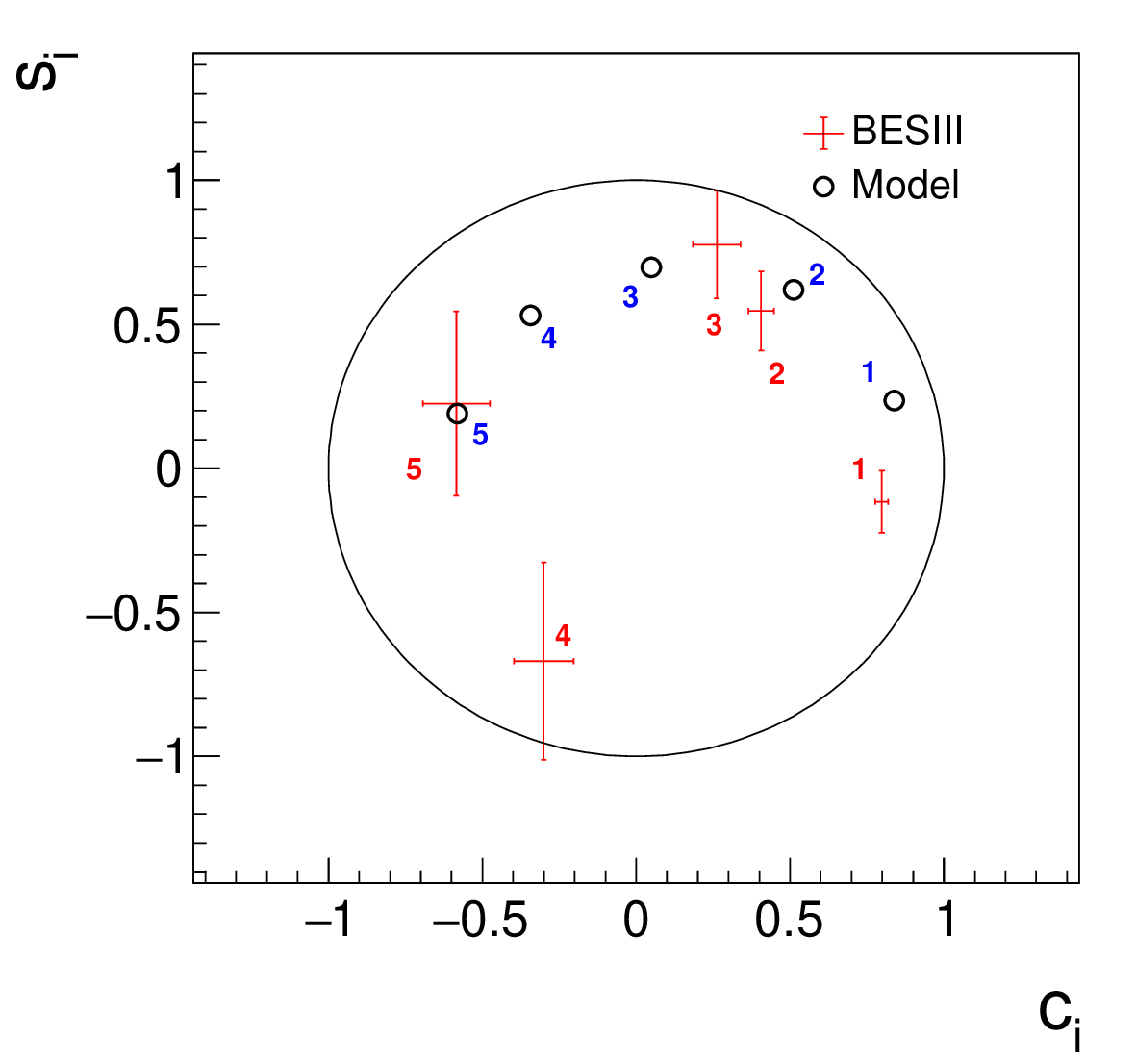}}   
\hspace{0.05cm}
\subfloat{\label{fig:CiSi_Optim}\includegraphics[width=0.5\textwidth]{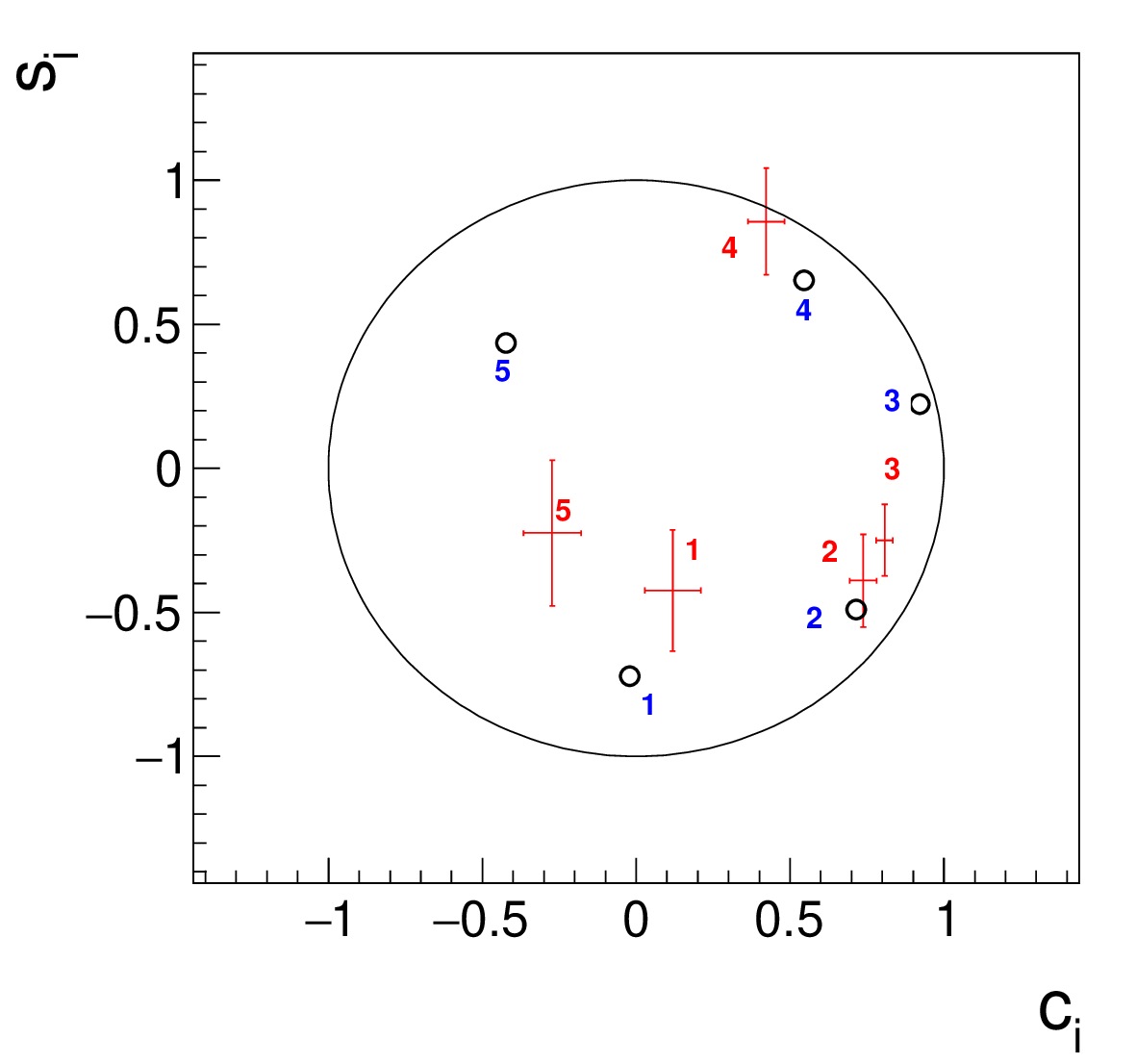}}
\caption[]{Results of the $c_i, s_i$ measurements for the equal-$\spp$-binning scheme (top) and the optimal-binning scheme (bottom). Also shown are the predictions based on the model from Ref.~\cite{BESIII:2023exz}.}
\label{fig:CiSi}
\end{figure}

\subsection{Determination of the $C\!P$-even fraction $F_+^{4\pi}$}

The strong-phase parameters for each bin are combined to determine the $C\!P$-even fraction $F_+^{4\pi}$ using the following equation:
\begin{linenomath*}
\begin{equation}
    F_+^{4\pi} = \frac{1}{2} + \sum_{i=1}^5 c_i \sqrt{T_i T_{-i}} \, . 
    \label{eq:fplus}
\end{equation}
\end{linenomath*}
The two binning schemes return compatible results, with the most precise measurement of $F_+^{4\pi} = 0.746 \pm 0.010 \pm 0.004$ coming from the equal-$\spp$ scheme.  This determination agrees well with the result of the inclusive analysis reported in  Ref.~\cite{BESIII:2022wqs}, but with an uncertainty that is approximately $30\%$ smaller.
 
\section{Systematic uncertainties}
\label{sec:syst}

There are several sources of possible systematic bias in the strong-phase parameters for which uncertainties are assigned.  These uncertainties are summarized in Table~\ref{tab:Syst1} for the equal-$\sp$-binning scheme, and in Table~\ref{tab:Syst2} for the optimal-binning scheme as shown in the appendix.   Unless stated otherwise, all systematic uncertainties are determined by repeatedly modifying the input parameter in question according to a normal distribution with width given by its uncertainty and refitting the values of $c_i$ and $s_i$.  The width of the distribution of fitted results, over approximately 1000 variations, is assigned as the corresponding systematic uncertainty.  The source of each uncertainty is explained below.

The uncertainties on the $K_i$ parameters of $D^0 \to \fpi$ are given in Table~\ref{tab:Ti} in the appendix, and derive from the statistical uncertainty in the number of flavor tags and the correction from the $D \to K^0_S \pi^+\pi^-$ peaking background. The DCS corrections, which are necessary for the hadronic flavor tags, have associated uncertainties that arise from the knowledge of the hadronic parameters of the decays, as listed in Table~\ref{tab:ExternalInputs}.

There is an uncertainty on the results associated with the knowledge of the fractional flavor-tagged yields $K_i^{\ks\pi\pi}$ and $K_i^{\kl\pi\pi}$ and strong-phase parameters $c_i^{\ks\pi\pi}$, $s_i^{\ks\pi\pi}$, $c_i^{\kl\pi\pi}$ and $s_i^{\kl\pi\pi}$ of the $D \to K^0_S \pi^+\pi^-$ and $D \to K^0_L \pi^+\pi^-$ tags. The uncertainties on these inputs are taken from the BESIII measurements reported in Ref.~\cite{BESIII:2020khq}, which, in the case of the strong-phase parameters, have been combined with the earlier measurements made at CLEO-c~\cite{CLEO:2010iul}.   Other external sources of uncertainty arise from the knowledge of the $C\!P$-even fraction in $D^0 \to \pi^+\pi^-\pi^0$ decays~\cite{Malde:2015mha} and the number of $D\bar{D}$ pairs produced in the data sample~\cite{BESIII:2018iev}.  

The efficiency matrices describing the migration from true to reconstructed phase-space bin for both the $D \to \fpi$ decay (see Eq.~(\ref{eq:EffMatrix})) and this decay when reconstructed together with $D \to K^0_{S,L}\pi^+\pi^-$ tags are determined from the MC simulation.  The statistical uncertainty arising from the finite MC sample size is propagated as a systematic uncertainty to the final results.

As can be seen in Eqs.~\eqref{eq:cptagwcorr} and~\eqref{eq:k0pipiwcor}, the ST yields presented in Table~\ref{tab:STDTYields} are used as input in the fit to calculate the expected DT yields for a given set of strong-phase parameters.
The total uncertainty on the ST yields comes from their statistical uncertainties and additional systematic contributions. To evaluate any bias associated with the fit procedure in determining these yields, alternative fits are performed on the $m_{\rm BC}$ distributions for ST samples, with variations of $\pm 0.5$~MeV applied to the end point of the ARGUS function. The contributions of the peaking backgrounds in all samples are varied according to the uncertainties of their branching fractions~\cite{pdg}. The variation in results is taken as the uncertainty associated with these sources. 
For the partially reconstructed channels, the uncertainty on the effective ST yield arises from the uncertainty in the factors given in Eq.~(\ref{eq:KlST}). 

The possible systematic bias associated with the determination of the DT yields is assessed following the same procedure as reported in Ref.~\cite{BESIII:2022wqs}. For the fully reconstructed tags, the end point of the ARGUS function is allowed to vary, similar to the ST study. For the partially reconstructed tags, the sizes of the peaking backgrounds are varied according to the uncertainties of their branching fractions. The corresponding changes in the $c_i$ and $s_i$ parameters are assigned as systematic uncertainties.

The total systematic uncertainties are obtained by summing the individual contributions in quadrature, assuming them to be uncorrelated. For the $c_i$ results, the dominant systematic uncertainty arises from the precision of the efficiency matrices, whereas for $s_i$, it derives from the knowledge of the strong-phase parameters of the $D \to K^0_{S,L}\pi^+\pi^-$ tags. In all cases the total systematic uncertainties are smaller than the statistical uncertainties.   

\begin{table*}[htbp]
\renewcommand\arraystretch{1.5}
  \centering
  \caption{Summary of systematic uncertainties for the equal-$\sp$-binning scheme.}
  \begin{tabular}{l c c c c c c c c c c }
  \toprule
	Source & $c_1$ & $c_2$ & $c_3$ & $c_4$ & $c_5$ & $s_1$ & $s_2$ & $s_3$ & $s_4$ & $s_5$  \\
  \hline   
  $K_i$ & 0.001 & 0.002 & 0.002 & 0.006 & 0.008 & 0.004 & 0.001 & 0.004 & 0.017 & 0.018 \\   
  $K_i^{\ks/\kl\pi\pi}$ & 0.001 & 0.002 & 0.003 & 0.002 & 0.003 & 0.003 & 0.010 & 0.008 & 0.011 & 0.012 \\  
  $c_i/s_i^{\ks/\kl\pi\pi}$ & 0.002 & 0.001 & 0.006 & 0.004 & 0.005 & 0.028 & 0.035 & 0.049 & 0.097 & 0.090 \\  
  $F_+^{\pi\pi\pi^0}$ & 0.005 & 0.002 & 0.001 & 0.001 & 0.003 & 0.001 & 0.000 & 0.000 & 0.000 & 0.000 \\  
  $N_{D\bar{D}}$ & 0.001 & 0.001 & 0.001 & 0.001 & 0.002 & 0.000 & 0.000 & 0.000 & 0.001 & 0.000 \\ 
  $\epsilon$ matrices & 0.002 & 0.005 & 0.011 & 0.029 & 0.025 & 0.001 & 0.001 & 0.001 & 0.004 & 0.004 \\
  ST yields & 0.004 & 0.007 & 0.008 & 0.007 & 0.008 & 0.000 & 0.001 & 0.001 & 0.001 & 0.001 \\  
  Peaking background & 0.004 & 0.000 & 0.006 & 0.000 & 0.002 & 0.000 & 0.000 & 0.000 & 0.000 & 0.000 \\ 
  DT yields & 0.001 & 0.001 & 0.007 & 0.005 & 0.001 & 0.001 & 0.001 & 0.001 & 0.001 & 0.001 \\ 
  \hline
  Total systematic & 0.008 & 0.010 & 0.017 & 0.031 & 0.029 & 0.029 & 0.036 & 0.050 & 0.095 & 0.092 \\
  Statistical & 0.022 & 0.041 & 0.077 & 0.097 & 0.109 & 0.108 & 0.137 & 0.187 & 0.343 & 0.321 \\  
  \bottomrule
  \end{tabular}
  \label{tab:Syst1}
\end{table*}

\begin{table*}[htbp]
\renewcommand\arraystretch{1.5}
  \centering
  \caption{Summary of systematic uncertainties for the optimal-binning scheme.}

  \begin{tabular}{l c c c c c c c c c c }
  \toprule
	Source & $c_1$ & $c_2$ & $c_3$ & $c_4$ & $c_5$ & $s_1$ & $s_2$ & $s_3$ & $s_4$ & $s_5$  \\
  \hline   
  $K_i$ & 0.007 & 0.011 & 0.002 & 0.010 & 0.012 & 0.006 & 0.003 & 0.006 & 0.007 & 0.026 \\    
  $K_i^{\ks/\kl\pi\pi}$ & 0.004 & 0.002 & 0.001 & 0.003 & 0.004 & 0.013 & 0.015 & 0.007 & 0.014 & 0.017 \\  
  $c_i/s_i^{\ks/\kl\pi\pi}$ & 0.005 & 0.002 & 0.003 & 0.005 & 0.005 & 0.040 & 0.059 & 0.029 & 0.074 & 0.073 \\ 
  $F_+^{\pi\pi\pi^0}$ & 0.000 & 0.009 & 0.009 & 0.002 & 0.002 & 0.000 & 0.001 & 0.001 & 0.000 & 0.000 \\
  $N_{D\bar{D}}$ & 0.002 & 0.001 & 0.000 & 0.001 & 0.002 & 0.000 & 0.000 & 0.000 & 0.000 & 0.001 \\  
  $\epsilon$ matrices & 0.014 & 0.007 & 0.003 & 0.007 & 0.017 & 0.001 & 0.001 & 0.001 & 0.001 & 0.001 \\ 
  ST yields & 0.011 & 0.005 & 0.004 & 0.009 & 0.010 & 0.001 & 0.000 & 0.000 & 0.001 & 0.001 \\ 
  Peaking background & 0.003 & 0.000 & 0.003 & 0.000 & 0.008 & 0.000 & 0.000 & 0.000 & 0.000 & 0.000 \\  
  DT yields & 0.007 & 0.001 & 0.005 & 0.004 & 0.001 & 0.001 & 0.001 & 0.001 & 0.001 & 0.001 \\ 
  \hline 
  Total systematic & 0.021 & 0.017 & 0.012 & 0.017 & 0.025 & 0.043 & 0.058 & 0.030 & 0.074 & 0.081 \\
  Statistical & 0.091 & 0.044 & 0.027 & 0.059 & 0.094 & 0.210 & 0.161 & 0.124 & 0.186 & 0.252 \\  
  \bottomrule
  \end{tabular}
  \label{tab:Syst2}
\end{table*}
\section{Impact on the $\gamma$ determination}

The results of the measurement are used as input in a study of the model-independent determination of $\gamma$ through $B^-\to DK^-, D\to\fpi$ decays.  The purpose of the study is to understand what additional uncertainty in $\gamma$ is induced by the uncertainty in the measurement of the $c_i$ and $s_i$ parameters.

Simulated pseudoexperiments are performed in which the decay rates follow the distributions given by Eq.~(\ref{eq:BGamma}). The values of $c_i$, $s_i$, $T_i$, and $T_{-i}$ are taken from the central values of the current analysis. The values of $\gamma$, $r_B$, and $\delta_B$ are set to $67^\circ$, $0.0986$, and $128^\circ$, respectively, which are close to the current world averages for these parameters~\cite{PhysRevD.84.033005,Bona:2005vz}.
For each pseudoexperiment, the yield in each phase-space bin is varied according to a Poisson distribution, and a very large number of signal events ($4\times 10^{6}$) are generated in order to ensure that the statistical uncertainty in the fit result associated with the number of $B$-meson decays is negligible. The fitted values of $\gamma$, $r_B$, and  $\delta_B$ are determined by maximizing the likelihood constructed by the Poisson function, where the expected values for $c_i$ and $s_i$ are the central values from the measurement, smeared according to the covariance matrices.  A total of $10^4$ pseudoexperiments are generated and fitted.

\begin{figure}[!htp]
\begin{center}
\begin{overpic}[width=0.5\textwidth]{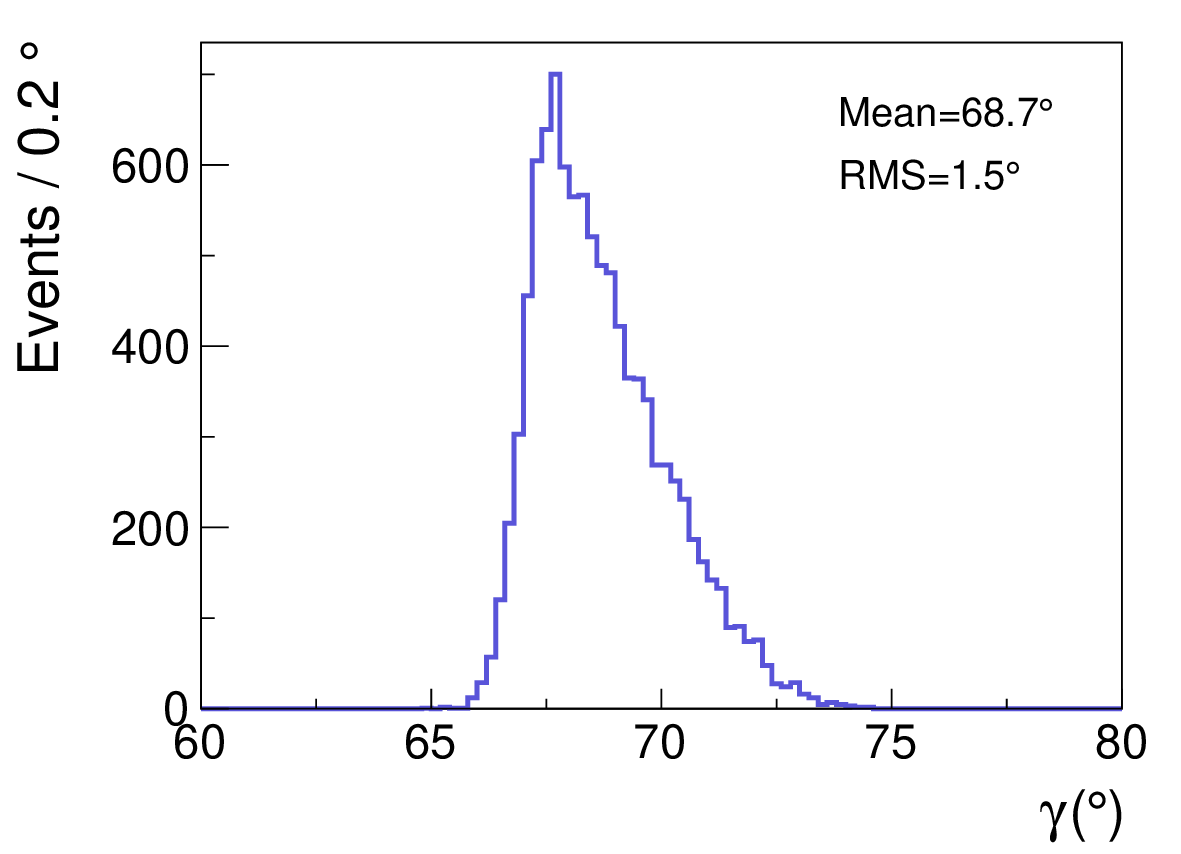}\put(25,60){}\end{overpic}
\begin{overpic}[width=0.5\textwidth]{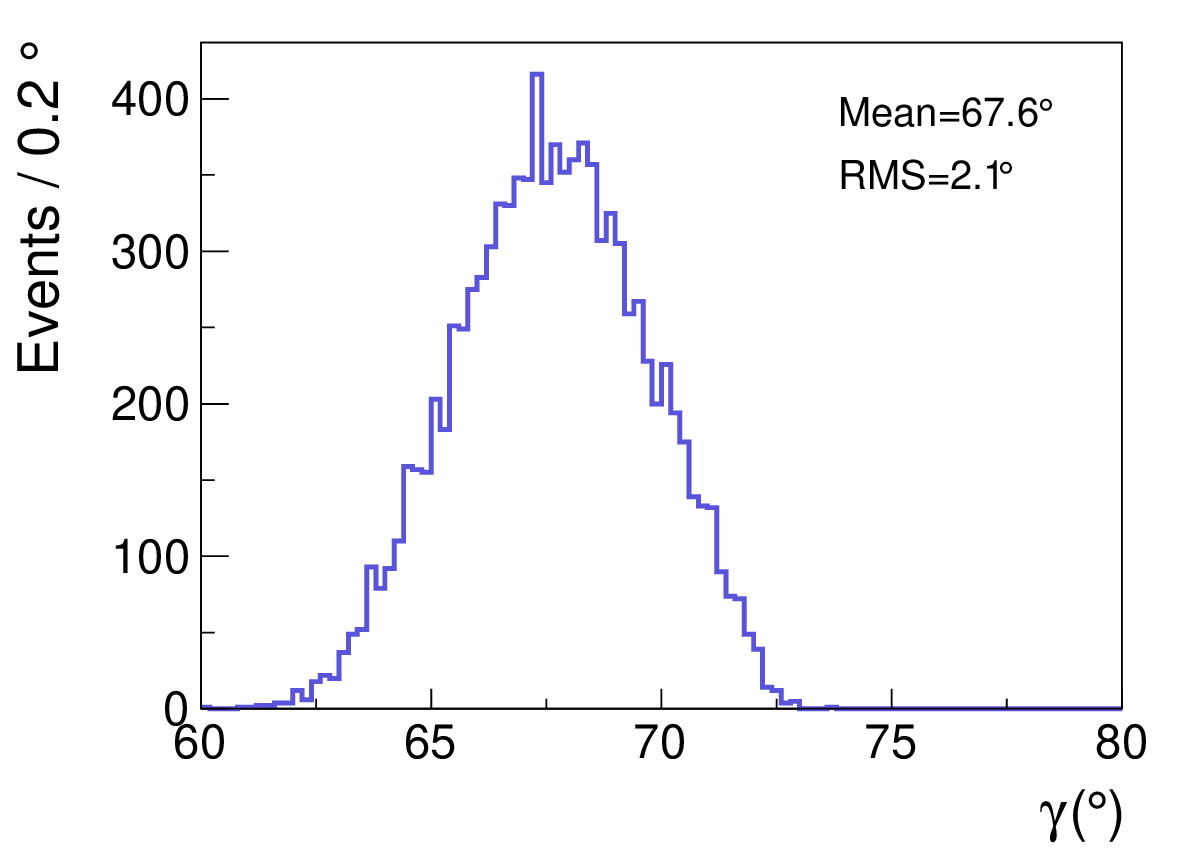}\put(25,60){}\end{overpic}
\caption[]{Distributions of fitted values of $\gamma$ for the equal-$\spp$-binning (top) and optimal-binning (bottom) schemes for very large samples of $B$-meson decays.}
\label{fig:Gamma}
\end{center}
\end{figure}

Figure~\ref{fig:Gamma} shows the distributions of the fitted values of $\gamma$ for both the equal-$\spp$-binning (top) and the optimal-binning (bottom) schemes. The distributions are not Gaussian, so the root mean square (RMS) is used to estimate the uncertainty on the $\gamma$ measurement arising from the knowledge of the $c_i$ and $s_i$ parameters.  The uncertainty for the equal-$\spp$-binning scheme is around $1.5^\circ$ and around $2^\circ$ for the optimal-binning scheme. These uncertainties apply in the limit of very high $B$-meson yields and increase for smaller samples. For example, for $10^4$ signal events, which is approximately the size of the currently available sample at LHCb~\cite{LHCb:2023yjo},
the uncertainties from the $c_i$ and $s_i$ parameters are  2.6$^\circ$ and 2.9$^\circ$ for the equal-$\spp$-binning and optimal-binning schemes, respectively. Both these numbers are smaller than the expected statistical uncertainty of such a measurement, which is found to be 6.6$^\circ$ for the equal-$\spp$-binning scheme and 5.8$^\circ$ for the optimal-binning scheme. Therefore, the optimal-binning scheme provides better overall precision, as expected, with the difference in performance in broad agreement with the behavior expected from the optimization metric discussed in Sec.~\ref{sec:formalism}.  
\section{Summary}

The strong-phase difference between $D^0$ and $\bar{D^0}$ decays to the final state $\pi^+\pi^-\pi^+\pi^-$ has been measured in bins of phase space, using a sample of quantum-correlated $D\bar{D}$ decays collected in $e^+e^-$ collisions at a c.m. energy of 3.773~GeV, corresponding to an integrated luminosity of 2.93~${\rm fb}^{-1}$.  The measurements were performed in two alternative schemes of $2 \times 5$ bins, both constructed from the amplitude model reported in Ref.~\cite{BESIII:2023exz}.   The equal-$\spp$ scheme divides the phase space into bins of equal strong-phase difference, whereas the optimal-binning scheme is designed to provide better sensitivity to the CKM angle $\gamma$ when used in an analysis of $B^\pm \to D K^\pm$ decays at LHCb or Belle II.  The measurements benefit from a larger sample of $D\bar{D}$ decays compared to a previous study~\cite{dArgent:2017gzv}, resulting in higher precision.  In the limit of a very large sample of  $B^\pm \to D K^\pm$ decays, the expected uncertainty on $\gamma$ arising from the knowledge of the strong-phase measurement is around $1.5^\circ$ for the equal-$\spp$-binning scheme and around $2^\circ$ for the optimal-binning scheme.  These uncertainties increase for smaller samples of $B$-meson decays, but remain lower than the statistical uncertainties arising from the sample size, with the optimal-binning scheme providing better overall precision when combining both statistical and systematic uncertainties. The measured strong-phase parameters in each bin are combined to give a value of $F_+^{4\pi} = 0.746 \pm 0.010 \pm 0.004$ for the $C\!P$-even fraction of the decay. This result is more precise than, and supersedes, the one reported in Ref.~\cite{BESIII:2022wqs}. $\textrm{BESIII}$ has now accumulated a larger data set of quantum-correlated $D\bar{D}$ pairs, corresponding to an integrated luminosity of 20~${\rm fb}^{-1}$, which will allow for more precise studies of the strong-phase differences in this and other decay modes.

\acknowledgments

The BESIII Collaboration thanks the staff of BEPCII and the IHEP computing center for their strong support. This work is supported in part by National Key R\&D Program of China under Contracts Nos. 2023YFA1606000, 2020YFA0406400, 2020YFA0406300; National Natural Science Foundation of China (NSFC) under Contracts Nos. 11635010, 11735014, 11935015, 11935016, 11935018, 12025502, 12035009, 12035013, 12061131003, 12192260, 12192261, 12192262, 12192263, 12192264, 12192265, 12221005, 12225509, 12235017, 12361141819; the Chinese Academy of Sciences (CAS) Large-Scale Scientific Facility Program; the CAS Center for Excellence in Particle Physics (CCEPP); Joint Large-Scale Scientific Facility Funds of the NSFC and CAS under Contract No. U1832207; 100 Talents Program of CAS; The Institute of Nuclear and Particle Physics (INPAC) and Shanghai Key Laboratory for Particle Physics and Cosmology; German Research Foundation DFG under Contracts Nos. FOR5327, GRK 2149; Istituto Nazionale di Fisica Nucleare, Italy; Knut and Alice Wallenberg Foundation under Contracts Nos. 2021.0174, 2021.0299; Ministry of Development of Turkey under Contract No. DPT2006K-120470; National Research Foundation of Korea under Contract No. NRF-2022R1A2C1092335; National Science and Technology fund of Mongolia; National Science Research and Innovation Fund (NSRF) via the Program Management Unit for Human Resources \& Institutional Development, Research and Innovation of Thailand under Contracts Nos. B16F640076, B50G670107; Polish National Science Centre under Contract No. 2019/35/O/ST2/02907; Swedish Research Council under Contract No. 2019.04595; The Swedish Foundation for International Cooperation in Research and Higher Education under Contract No. CH2018-7756; U. S. Department of Energy under Contract No. DE-FG02-05ER41374.

\bibliographystyle{apsrev4-1} 
\bibliography{references}

\section*{Appendix: correlation matrices for the $c_i$, $s_i$ measurements}
\label{app:Corr}
The correlation matrices for the statistical and systematic uncertainties of the $c_i$ and $s_i$ measurements are given in Tables~\ref{tab:EqualStatCorr}, \ref{tab:EqualSystCorr}, \ref{tab:OptimStatCorr} and \ref{tab:OptimSystCorr}.
\begin{table*}[!hbp]
\renewcommand\arraystretch{1.5}
\caption{The correlation matrix for the statistical uncertainties for the equal-$\sp$-binning scheme.}
\label{tab:EqualStatCorr}
\centering
\begin{tabular}{ c | c c c c c c c c c  }
  \toprule 
	   & $c_2$ & $c_3$ & $c_4$ & $c_5$ & $s_1$ & $s_2$ & $s_3$ & $s_4$ & $s_5$  \\
  \hline 
    $c_1$& $-$0.084 & $-$0.006 & $-$0.004 & $-$0.004 & $-$0.023 & $-$0.001 & 0 & 0 & 0 \\  
    $c_2$& & $-$0.096 & $-$0.006 & $-$0.004 & 0 & $\phantom{-}$0.020 & $-$0.003 & 0 & 0 \\  
    $c_3$&  &  & $-$0.110 & $-$0.020 & 0 & $-$0.003 & $\phantom{-}$0.025 & $\phantom{-}$0.001 & $-$0.001 \\  
    $c_4$& & &  & $-$0.139 & 0 & 0 & $\phantom{-}$0.001 & $-$0.021 & $-$0.002 \\  
    $c_5$& &  &  &  & 0 & 0 & 0 & $-$0.001 & $\phantom{-}$0.050 \\  
    $s_1$& &  &  &  &   & $-$0.071 & $-$0.001 & $-$0.001 & $-$0.001 \\  
    $s_2$&  & &  &  &  & & $-$0.091 & $-$0.005 & $-$0.002 \\  
    $s_3$& & & &  &  &  &  & $-$0.107 & $-$0.006 \\  
    $s_4$&  &  & &  & &  &  &   & $-$0.086 \\  
  \bottomrule
\end{tabular}
\end{table*}

\begin{table*}[!hbp]
\renewcommand\arraystretch{1.5}
\caption{The correlation matrix for the statistical uncertainties for the optimal-binning scheme.}
\label{tab:OptimStatCorr}
\centering
\begin{tabular}{ c | c c c c c c c c c  }
  \toprule 
	  &   $c_2$ & $c_3$ & $c_4$ & $c_5$ & $s_1$ & $s_2$ & $s_3$ & $s_4$ & $s_5$  \\
  \hline 
 $c_1$&  $-$0.057 & $-$0.001 & $-$0.026 & $-$0.073 & $-$0.036 & $\phantom{-}$0.001 & 0 & 0 & $-$0.001 \\  
 $c_2$&  & $-$0.061 & $-$0.030 & $-$0.001 & $\phantom{-}$0.002 & $-$0.016 & $\phantom{-}$0.001 & $\phantom{-}$0.001 & 0 \\  
 $c_3$& &  &  $-$0.040 & 0 & 0 & $\phantom{-}$0.001 & $-$0.016 & $\phantom{-}$0.002 & 0 \\  
 $c_4$&  &  &  &  $-$0.048 & 0 & 0 & $\phantom{-}$0.001 & $-$0.035 & 0 \\  
 $c_5$&  &  &  &  &  0.002 & 0 & 0 & 0 & $\phantom{-}$0.018 \\  
 $s_1$&  & &  &  &  &  $-$0.054 & 0 & $\phantom{-}$0.004 & $\phantom{-}$0.001 \\  
 $s_2$&  & & &  &  & &  $\phantom{-}$0.016 & $\phantom{-}$0.009 & $-$0.001 \\  
 $s_3$&  &  &  &  &  &  &  &  $-$0.037 & 0 \\  
 $s_4$&  & &  &  &  &  &  &  &  $-$0.044 \\  
  \bottomrule 
\end{tabular}
\end{table*}

\begin{table*}[!hbp]
\renewcommand\arraystretch{1.5}
\caption{The correlation matrix for the systematic uncertainties for the equal-$\sp$-binning scheme.}
\label{tab:EqualSystCorr}
\centering
\begin{tabular}{ c | c c c c c c c c c  }
  \toprule 
	  &  $c_2$ & $c_3$ & $c_4$ & $c_5$ & $s_1$ & $s_2$ & $s_3$ & $s_4$ & $s_5$  \\
  \hline 
 $c_1$& 0.640 & 0.411 & $\phantom{-}$0.149 & $\phantom{-}$0.115 & $-$0.072 & $\phantom{-}$0.029 & $-$0.014 & $-$0.024 & 0.043 \\  
 $c_2$&  & 0.346 & $\phantom{-}$0.162 & $\phantom{-}$0.133 & 0.011 & $\phantom{-}$0.108 & $\phantom{-}$0.023 & $-$0.033 & 0.050 \\  
 $c_3$&  &  & $-$0.047 & $\phantom{-}$0.190 & 0.084 & $\phantom{-}$0.062 & $\phantom{-}$0.116 & $\phantom{-}$0.085 & 0.035 \\  
 $c_4$&  &  &  & $-$0.072 & $-$0.048 & $-$0.080 & $-$0.011 & $\phantom{-}$0.046 & 0.026 \\  
 $c_5$&  & &  &  & $-$0.012 & $\phantom{-}$0.061 & $\phantom{-}$0.069 & $-$0.021 & 0.040 \\  
 $s_1$&  &  &  &  &  & $-$0.008 & $-$0.295 & $\phantom{-}$0.409 & 0.173 \\  
 $s_2$&  &  &  &  &  &  & $\phantom{-}$0.601 & $-$0.474 & 0.100 \\  
 $s_3$&  &  &  &  &  &  &  & $-$0.241 & 0.035 \\     
 $s_4$&  &  &  &  &  &  &  &   & 0.101 \\    
  \bottomrule 
\end{tabular}
\end{table*}

\begin{table*}[!hbp]
\renewcommand\arraystretch{1.5}
\caption{The correlation matrix for the systematic uncertainties for the optimal-binning scheme.}
\label{tab:OptimSystCorr}
\centering
\begin{tabular}{ c | c c c c c c c c c  }
  \toprule 
	   & $c_2$ & $c_3$ & $c_4$ & $c_5$ & $s_1$ & $s_2$ & $s_3$ & $s_4$ & $s_5$  \\
  \hline 
 $c_1$ & 0.181 & 0.318 & 0.321 & 0.26 & $-$0.056 & $-$0.029 & $-$0.082 & $-$0.017 & $\phantom{-}$0.012 \\  
 $c_2$&  & 0.566 & 0.244 & 0.099 & $-$0.027 & $-$0.097 & $-$0.081 & $-$0.028 & $-$0.031 \\  
 $c_3$&  &  & 0.363 & 0.241 & $\phantom{-}$0.014 & $\phantom{-}$0.002 & $-$0.105 & $\phantom{-}$0.011 & $-$0.050 \\  
 $c_4$&  &  &  & 0.268 & $-$0.152 & $-$0.006 & $\phantom{-}$0.153 & $\phantom{-}$0.058 & $\phantom{-}$0.042 \\  
 $c_5$&  &  &  &  & $-$0.005 & $\phantom{-}$0.034 & $-$0.015 & $\phantom{-}$0.009 & $\phantom{-}$0.142 \\  
 $s_1$&  &  &  &  &  & $\phantom{-}$0.313 & $\phantom{-}$0.061 & $\phantom{-}$0.064 & $\phantom{-}$0.170 \\  
 $s_2$&  &  &  &  &  &  & $\phantom{-}$0.476 & $\phantom{-}$0.075 & $\phantom{-}$0.328 \\  
 $s_3$&  &  &  &  &  &  &  & $\phantom{-}$0.102 & $\phantom{-}$0.455 \\  
 $s_4$&  &  &  &  &  &  &  &  & $\phantom{-}$0.134 \\  
  \bottomrule 
\end{tabular}
\end{table*}


\end{document}